\documentclass[aps]{revtex4}
\usepackage{amsfonts}
\usepackage{amsmath}
\usepackage{amssymb,epsf}
\usepackage{graphicx}

\begin{document}

\title{Three dimensional magnetic solutions in massive gravity with
(non)linear field}
\author{S. H. Hendi$^{1,2}$\footnote{\emph{Present address:}
hendi@shirazu.ac.ir}, B. Eslam
Panah$^{1,2,3}$\footnote{\emph{Present address:}
behzad.eslampanah@gmail.com}, S. Panahiyan$^{1,4,5}$
\footnote{\emph{Present
address:} shahram.panahiyan@uni-jena.de} and M. Momennia$^{1}$\footnote{\emph{%
Present address:} m.momennia@shirazu.ac.ir}}
\affiliation{$^1$Physics Department and Biruni Observatory, College of Sciences, Shiraz
University, Shiraz 71454, Iran\\
$^2$Research Institute for Astrophysics and Astronomy of Maragha (RIAAM),
P.O. Box 55134-441, Maragha, Iran\\
$^3$ICRANet, Piazza della Repubblica 10, I-65122 Pescara, Italy\\
$^4$Helmholtz-Institut Jena, Fr\"{o}belstieg 3, Jena D-07743, Germany \\
$^5$Physics Department, Shahid Beheshti University, Tehran 19839, Iran }

\begin{abstract}
The Noble Prize in physics 2016 motivates one to study different aspects of
topological properties and topological defects as their related objects.
Considering the significant role of the topological defects (especially
magnetic strings) in cosmology, here, we will investigate three dimensional
horizonless magnetic solutions in the presence of two generalizations:
massive gravity and nonlinear electromagnetic field. The effects of these
two generalizations on properties of the solutions and their geometrical
structure are investigated. The differences between de Sitter and anti de
Sitter solutions are highlighted and conditions regarding the existence of
phase transition in geometrical structure of the solutions are studied.
\end{abstract}

\maketitle

\section{Introduction}

The Nobel Prize in physics $2016$ has been assigned to the interesting
consequences of topological invariant and topological phase transitions.
Although general relativity is based on the local transformation, it is
found that topological properties have significant impacts on our
understanding of the universe. In this regard, topological defects have been
observed in the various branches of physics. The crucial role of topological
defects was observed in a new type of phase transition in two-dimensional
systems \cite{KosterlitzThouless1}. Kosterlitz and Thouless applied the
mentioned points to superconducting and superfluid films and they found
their important roles in quantum nature of one-dimensional systems at very
low temperatures. In addition, it was shown that phenomenological properties
of different phases of physical systems could be explained by these
topological defects. For example, in studying liquid crystal, it was shown
that structural properties and phase transitions are affected by topological
effects \cite{liquid1}. The applications of these defects in condensed
matter with ordered media \cite{media}, magnetism and nanomagnetism \cite%
{magnetism1}, vortices in superfluid \cite{superfluids} and Bose-Einstein
condensate \cite{bose1} were explored. Furthermore, the importance of these
mathematical tools in studying superconductors and their phase transitions
were highlighted in Refs. \cite{super1,super2}.

From the cosmological point of view, existence of the topological defects
could be traced back into early universe and also phase transitions in the
early universe \cite{Kibble1}. The existence of topological defects is
originated from breaking down the symmetry in phase transitions that has
taken place in the early universe. Speaking more precisely, regions of the
universe which are separated by more than the distance $d=ct$ (in which $c$
is speed of light and $t$ is time), could not know anything about each
other. During the phase transition, different regions choose different
minima in the set of possible states to fall in. The topological defects are
placed at the boundaries of these regions which have chosen different
minima. Therefore, one can state that topological defects are the results of
disagreement between different choices of different regions. In the context
of cosmology, there are different types of the topological defects.
Depending on the dimensionality and structural properties, these defects
could be categorized into; (i) Domain walls which are due to a broken
discrete symmetry and divide universe into blocks. (ii) Cosmic strings which
are due axial or cylindrical symmetry breaking and are related to grand
unified particle physics models/electroweak scale. (iii) Monopoles which are
super massive and carry magnetic charge and are formed when a spherical
symmetry is broken. (iv) Textures which are formed due to the breaking of
several symmetries. These topological defects carry information regarding
the early universe. In addition, it was proposed that they could have
specific roles in the large-scale structures \cite{Kibble1}, anisotropy in
the Cosmic Microwave Background (CMB) \cite{AC1} and dark matter \cite{dark1}%
. Besides, these topological defects could be used as cosmological lenses
\cite{lens}. In other words, the trajectory of the photon on these
topological defects are affected depending on deficit angle. This highlights
the importance of analyzing deficit angle in the properties of topological
defects.

The cosmic strings in the presence of Maxwell field have been investigated
\cite{OJCDias1}. Furthermore, the superconducting property of these
topological defects has been explored in Einstein \cite{Witten1}, dilaton
\cite{CNFerreira1} and Brans-Dicke \cite{AASen} theories. In addition, the
QCD applications of the magnetic strings \cite{QCD1} and their roles in
quantum theories \cite{Quantum1} have been investigated before. The
stability of the cosmic strings through quantum fluctuations has been
analyzed in Ref. \cite{stable}. The limits on the cosmic string tension have
been studied by extracting signals of cosmic strings from CMB temperature
anisotropy maps \cite{CMB}. The spectrum of gravitational wave background
produced by cosmic strings is obtained in Ref. \cite{spectrum}. For further
investigations regarding cosmic strings, we refer the reader to an
incomplete list of references \cite{cosmic1}.

Domain walls and their evolution in de Sitter universe have been studied in
\cite{evolution}. In addition, the gravitational waves produced from
decaying domain walls are investigated in Ref. \cite{gravwave}. The
localization of the fields on the dynamical domain wall was investigated and
it was shown that the chiral spinor can be localized on the domain walls
\cite{local}. For further studies regarding this class of topological
defects, we refer the reader to Ref. \cite{Skenderis1}.

On the other hand, considering most of physical systems in nature, one finds
that they exhibit nonlinear behavior, and therefore, the nonlinear field
theories are of importance in physical researches. There are many
motivations for studying the nonlinear electrodynamics (NED) such as; (i)
These theories are the generalizations of Maxwell field and reduce to linear
Maxwell theory in the special cases (weak nonlinearity). (ii) These
nonlinear theories can describe the radiation propagation inside specific
materials \cite{De Lorenci1}. (iii) Some special NED models can describe the
self-interaction of virtual electron-positron pairs \cite{Heisenberg}. (iv)
Theories of NED can remove the problem of point-like charge self-energy. (v)
From the standpoint of quantum gravity and its coupling with these nonlinear
theories, we can obtain more information and deep insight regarding the
nature of gravity \cite{Seiberg}. (vi) Compatibility with AdS/CFT
correspondence and string theory are other properties of NED theories. (vii)
NED theory improves the basic concept of gravitational redshift and its
dependency of any background magnetic field as compared to the
well-established method introduced by general relativity. (viii) From the
perspective of cosmology, it was shown that NED theories can remove both of
the big bang and black hole singularities \cite{Ayon2}. (ix) From
astrophysical point of view, it was found that the effects of NED become
indeed quite important in super-strong magnetized compact objects, such as
pulsars and particular neutron stars \cite{Bialynicka}.

There are different models of NED, such as Born-Infeld form \cite{BornI},
logarithmic form \cite{Soleng}, exponential form \cite{HendiJHEP},
arcsin-electrodynamics form \cite{KruglovI} and etc. One of the interesting
branches of the nonlinear electrodynamics is power Maxwell invariant (PMI)
theory. The Lagrangian of PMI theory is an arbitrary power of the Maxwell
Lagrangian \cite{Hassaine1} which could reduce to the Maxwell field by
choosing the unit power. In addition, the PMI theory has an interesting
consequence which distinguishes this NED theory from other theories; this
theory enjoys conformal invariancy when the power of Maxwell invariant is a
quarter of space-time dimensions (power = dimensions/4). In other words, in
this case, the energy-momentum tensor will be a traceless tensor which leads
to conformal invariancy and also an inverse square law of the electric field
for the point-like charge in arbitrary dimensions \cite{Hassaine1}.

Recent observations of gravitational waves from a binary black hole merger
in LIGO and Virgo collaboration provided a deep insight to general
relativity (GR) and existence of massive gravitons \cite{Abbott}. However,
gravitons are massless particles with spin $2$ in GR which have two degrees
of freedom. Since the quantum theory of massless gravitons is
non-renormalizable \cite{Deser}, in order to remove this problem, one may
modify GR to massive gravity by adding a mass term to the Einstein-Hilbert
action. Therefore, considering this action, the graviton will have a mass of
$m$ which in case of $m\rightarrow 0$, the effect of massive gravity will be
vanished. In other words, massive gravity is a modification of GR that
gravitons have mass. Among the motivations of the massive gravity one can
mention description of accelerating expansion of universe without
considering the cosmological constant \cite{Deffayet2}. It was shown that
the terms of massive gravitons can be equivalent to a cosmological constant
\cite{Gumrukcuoglu1}. This theory modifies gravity compared with GR which
allows the universe to accelerate at the large scale, however at small
scale, this theory reduces to GR as well. This theory of gravity may
illustrate the dark energy problem \cite{Dvali1}. In addition, the existence
of massive gravitons provides extra polarization for gravitational waves,
and affects the propagation's speed of the gravitational waves \cite{Will},
hence, the production of gravitational waves during inflation \cite{Mohseni}%
. By adding the interaction terms to GR, massive gravity with flat
background was investigated by Fierz and Pauli \cite{Fierz}. However, this
theory suffers a van Dam-Veltman-Zakharov (vDVZ) discontinuity \cite{vDV}.
Generalization of massive theory to curved background was done by
Boulware-Deser. This generalization leads to the existence of a typical
ghost, the so-called Boulware-Deser ghost \cite{BoulwareD}. Several models
of massive theory were proposed by some authors in order to avoid
discontinuity and ghost problems \cite{Ohta}. One of the ghost-free massive
theories in three dimensions was introduced by Bergshoeff, Hohm and Townsend
(new massive gravity (NMG)) \cite{BergshoeffHT}. However, NMG has ghost
problem in four and higher dimensions. Therefore, in order to resolve ghost
problem in diverse dimensions, a new theory of massive gravity was proposed
by de Rham, Gabadadze and Tolley (dRGT) in 2011 \cite{de Rham1}. The
stability of dRGT massive theory was studied and it was shown that such
theory enjoys absence of the Boulware-Deser ghost \cite{Hassan1}. Black hole
and cosmological solutions have been investigated in dRGT massive gravity
\cite{Fasiello,Babichev1,Bamba,CaiS,Goon,Solomon,Pan,LiLX,Cao}. Also,
reentrant phase transitions of higher-dimensional AdS black holes and
behavior of quasinormal modes and van der Waals like phase transition of
charged AdS black holes in massive gravity have been studied in Refs. \cite%
{Zou1,Zou2}.

It is notable that in massive gravity theory, the mass terms are produced by
consideration of a reference metric. Considering the reference metric in
massive gravity, one finds that it plays a crucial role in construction of
exact solutions \cite{de Rham3}. In this regard, Vegh introduced a new
reference metric which was motivated by applications of gauge/gravity
duality \cite{Vegh}. It is believed that the graviton may behave like a
lattice and exhibits a Drude peak in this model of massive theory \cite{Vegh}%
. Another property of this model is related to ghost-free and stability for
arbitrary singular metric \cite{Zhang}. The action of massive gravity in an
arbitrary $d-$dimensions is given by
\begin{equation}
\mathcal{I}_{G}=-\frac{1}{16\pi }\int_{\mathcal{M}}d^{d}x\sqrt{-g}\left[
\mathcal{R}+m^{2}\sum_{i=1}^{4}c_{i}\mathcal{U}_{i}(g,f)\right] ,
\label{action}
\end{equation}%
where $\mathcal{R}$\ is the scalar curvature and $m^{2}$ is related to the
mass of gravitons. In addition, $f$ is a fixed symmetric tensor, $c_{i}$'s
are some constants, and $\mathcal{U}_{i}$'s are symmetric polynomials of the
eigenvalues of matrix $\mathcal{K}_{\nu }^{\mu }=\sqrt{g^{\mu \alpha
}f_{\alpha \nu }}$ which are as follow
\begin{eqnarray}
\mathcal{U}_{1} &=&\left[ \mathcal{K}\right] ,\;\;\;\;\;\mathcal{U}_{2}=%
\left[ \mathcal{K}\right] ^{2}-\left[ \mathcal{K}^{2}\right] ,\;\;\;\;\;%
\mathcal{U}_{3}=\left[ \mathcal{K}\right] ^{3}-3\left[ \mathcal{K}\right] %
\left[ \mathcal{K}^{2}\right] +2\left[ \mathcal{K}^{3}\right] ,  \notag \\
&&\mathcal{U}_{4}=\left[ \mathcal{K}\right] ^{4}-6\left[ \mathcal{K}^{2}%
\right] \left[ \mathcal{K}\right] ^{2}+8\left[ \mathcal{K}^{3}\right] \left[
\mathcal{K}\right] +3\left[ \mathcal{K}^{2}\right] ^{2}-6\left[ \mathcal{K}%
^{4}\right] .  \notag
\end{eqnarray}

Charged black hole solutions with (non)linear field and the existence of van
der Waals like behavior in extended phase space and also geometrical
thermodynamics by considering dRGT massive gravity have been studied \cite%
{CaiMassive,HendiEP1,HendiPEM}. Moreover, the hydrostatic equilibrium
equation of neutron stars by using this theory of massive gravity was
obtained and it was shown that the maximum mass of neutron stars can be
about $3.8M_{\odot }$ (where $M_{\odot }$ is mass of the sun) \cite{HendiBEP}%
. Also, holographic conductivity in this gravity with PMI field has been
investigated in Ref. \cite{Dehyadegari}. Besides, the generalization of this
theory to include higher derivative gravity \cite{HendiPE} and gravity's
rainbow \cite{HendiEP2} has been done in literatures. In addition, three
dimensional (BTZ) charged black hole solutions with (non)linear field have
been studied in Ref. \cite{HendiEP3}.

By adding an electromagnetic Lagrangian ($\mathcal{L}(\mathcal{F})$) and the
cosmological constant ($\Lambda $) to the action (\ref{action}) with $d=3$,
we have
\begin{equation}
\mathcal{I}_{G}=-\frac{1}{16\pi }\int_{\mathcal{M}}d^{3}x\sqrt{-g}\left[
\mathcal{R}-2\Lambda +\mathcal{L}(\mathcal{F})+m^{2}\sum_{i=1}^{4}c_{i}%
\mathcal{U}_{i}(g,f)\right] .  \label{Action}
\end{equation}

Varying the action (\ref{Action}) with respect to the gravitational and
gauge fields, one can obtain the following field equations
\begin{equation}
R_{\mu \nu }-\frac{1}{2}g_{\mu \nu }\left( \mathcal{R}-2\Lambda \right)
+m^{2}\chi _{\mu \nu }=T_{\mu \nu },  \label{Field equation}
\end{equation}%
\begin{equation}
\partial _{\mu }\left( \sqrt{-g}\mathcal{L}_{\mathcal{F}}F^{\mu \nu }\right)
=0,  \label{Maxwell equation}
\end{equation}%
in which $\mathcal{L}_{\mathcal{F}}=d\mathcal{L}(\mathcal{F})/d\mathcal{F}$
where $\mathcal{F}=F_{\mu \nu }F^{\mu \nu }$\ is the Maxwell invariant, $%
F_{\mu \nu }$\ $=\partial _{\mu }A_{\nu }-\partial _{\nu }A_{\mu }$ is the
Faraday tensor and $A_{\mu }$ is the gauge potential. In addition, $\chi
_{\mu \nu }$ is the massive term with the following form
\begin{eqnarray}
\chi _{\mu \nu } &=&-\frac{c_{1}}{2}\left( \mathcal{U}_{1}g_{\mu \nu }-%
\mathcal{K}_{\mu \nu }\right) -\frac{c_{2}}{2}\left( \mathcal{U}_{2}g_{\mu
\nu }-2\mathcal{U}_{1}\mathcal{K}_{\mu \nu }+2\mathcal{K}_{\mu \nu
}^{2}\right) -\frac{c_{3}}{2}(\mathcal{U}_{3}g_{\mu \nu }-3\mathcal{U}_{2}%
\mathcal{K}_{\mu \nu }+  \notag \\
&&6\mathcal{U}_{1}\mathcal{K}_{\mu \nu }^{2}-6\mathcal{K}_{\mu \nu }^{3})-%
\frac{c_{4}}{2}(\mathcal{U}_{4}g_{\mu \nu }-4\mathcal{U}_{3}\mathcal{K}_{\mu
\nu }+12\mathcal{U}_{2}\mathcal{K}_{\mu \nu }^{2}-24\mathcal{U}_{1}\mathcal{K%
}_{\mu \nu }^{3}+24\mathcal{K}_{\mu \nu }^{4}),  \label{massiveTerm}
\end{eqnarray}%
and the energy-momentum tensor of Eq. (\ref{Field equation}) is
\begin{equation}
T_{\mu \nu }=\frac{1}{2}g_{\mu \nu }\mathcal{L}(\mathcal{F})-2\mathcal{L}_{%
\mathcal{F}}F_{\mu \lambda }F_{\nu }^{\lambda }.  \label{Energy momentum}
\end{equation}

Here, we want to obtain the magnetic solutions of Eqs. (\ref{Field
equation}) and (\ref{Maxwell equation}) by considering the Maxwell
electromagnetic field ($\mathcal{L}(\mathcal{F})=-\mathcal{F}$).

Magnetic branes (or horizonless solution) are interesting objects
which have been investigated by many authors
\cite{Mag1,Mag2,Mag3,Mag4,Mag5,Mag7,Mag12,ThreeDim}. Our main
motivation here is to understand the effects of two
generalizations on the magnetic horizonless solutions with
interpretation of topological defects. These two generalizations
include massive gravity and PMI electromagnetic field. Considering
the applications of topological defects in dark matter, CMB,
gravitational waves, large scale structure and etc., it is
necessary to investigate the effects of the massive gravitons on
the structure and formation of topological defects. Here, we
intend to show how generalization to massive gravity would modify
geometrical structure of the magnetic solutions. To do so, we
apply the massive gravity generalization and investigate
geometrical properties such as deficit angle. Considering the
electromagnetically charged aspect of the objects of interest in
this paper (magnetic solutions), we will take two cases of linear
and nonlinear electromagnetic fields into account. Here, we would
investigate the effects of Maxwell and PMI electromagnetic fields
on the deficit angle, hence, geometrical structure of the
topological defects known as horizonless magnetic solutions. The
combinations of massive gravity and PMI theory is another subject
of interest which would be addressed. It is notable that such
magnetic source was interpreted as a kind of magnetic monopole
reminiscent of a Nielson-Oleson vortex solution \cite{Hirschmann},
while Dias and Lemos interpreted it as a composition of two
symmetric and superposed electric charges \cite{OJCDias1}. In
other words, one of the mentioned electric charges is at rest and
the other is rotating, and therefore, there is no electric field
since the total electric charge is zero, but angular electric
current produces a magnetic field.

Now, we use the new metric of three dimensional spacetime with
$(-++)$ signature which was introduced in Ref. \cite{ThreeDim}
\begin{equation}
ds^{2}=-\frac{\rho ^{2}}{l^{2}}dt^{2}+\frac{d\rho ^{2}}{g(\rho )}%
+l^{2}g(\rho )d\varphi ^{2},  \label{metric}
\end{equation}%
where $g(\rho )$ is an arbitrary function of radial coordinate
$\rho $ which should be determined. The scale length factor $l$ is
related to the cosmological constant $\Lambda $, and the angular
coordinate $\varphi $ is dimensionless as usual and ranges in
$[0,2\pi ]$. The motivation of considering the metric gauge
[$g_{tt}\propto -\rho ^{2}$ and $\left( g_{\rho \rho }\right)
^{-1}\propto g_{\varphi \varphi }$] instead of the usual
Schwarzschild like gauge [$\left( g_{\rho \rho }\right)
^{-1}\propto g_{tt}$ and $g_{\varphi \varphi }\propto \rho ^{2}$]
comes from the fact that we are looking for magnetic solutions
without curvature singularity. It is easy to show that using a
suitable transformation, the metric (\ref{metric}) can be mapped
to $3$-dimensional Schwarzschild like spacetime locally, but not
globally \cite{ThreeDim}.

In order to obtain exact solutions, we should make a choice for the
reference metric. We consider the following ansatz metric
\begin{equation}
f_{\mu \nu }=diag(\frac{-c^{2}}{l^{2}},0,0),  \label{f11}
\end{equation}%
where in the above equation $c$ is a positive constant. Using the metric
ansatz (\ref{f11}), $\mathcal{U}_{i}$'s are \cite{CaiMassive,HendiEP3}
\begin{equation}
\mathcal{U}_{1}=\frac{c}{\rho },\;\;\;\;\;\mathcal{U}_{2}=\mathcal{U}_{3}=%
\mathcal{U}_{4}=0,  \label{U}
\end{equation}%
which indicate that the only contribution of massive gravity comes from $%
\mathcal{U}_{1}$ in three dimensions. \ Before proceeding we give a reason
for such choice of the reference metric (\ref{f11}). For three dimensional
black holes, the spacetime metric with $(-,+,+)$\ signature has the
following explicit form
\begin{equation}
ds^{2}=-g(\rho )dt^{2}+\frac{d\rho ^{2}}{g(\rho )}+\rho ^{2}d\varphi ^{2}.
\label{three}
\end{equation}

In order to obtain exact black hole solutions, we consider the ansatz metric
as $f_{\mu \nu }=diag(0,0,c^{2})$ (see Refs. \cite{Vegh}, \cite{CaiMassive}
and \cite{HendiEP1}, for more details). Here, the metric function ($g(\rho )$%
) is factors of radial and spatial coordinates in magnetic spacetime metric
(Eq. \ref{metric}). In order to have exact solutions in an axially symmetric
spacetime with the form (\ref{metric}),\ it is necessary to consider the
reference metric as $f_{\mu \nu }=diag(\frac{-c^{2}}{l^{2}},0,0)$. This form
of reference metric is expectable. Comparing black hole metric, Eq. (\ref%
{three}), with magnetic spacetime, Eq. (\ref{metric}), we find that Eq. (\ref%
{metric}) can be reproduced from Eq. (\ref{three}) by the following local
transformations:%
\begin{equation}
t\longrightarrow il\varphi \;\;\;\&{\ \ \ \ }\varphi \longrightarrow it/l.
\end{equation}%
Since we changed the role of $t$\ and $\varphi $\ coordinates, the nonzero
component of the reference metric should be changed accordingly.

Since we are going to study the linearly magnetic solutions, we choose the
Lagrangian of Maxwell field $\mathcal{L}(\mathcal{F})=-\mathcal{F}$ for Eqs.
(\ref{Action}), (\ref{Maxwell equation}), and (\ref{Energy momentum}). It is
well-known that the electric field is associated with the time component of
the vector potential $A_{t}$, while the magnetic field is associated with
the angular component $A_{\varphi }$. Due to our interest to investigate the
magnetic solutions, we assume the vector potential as
\begin{equation}
A_{\mu }=h(\rho )\delta _{\mu }^{\varphi }.  \label{Gauge potential}
\end{equation}

Using the Maxwell equation (\ref{Maxwell equation}) with $\mathcal{L}(%
\mathcal{F})=-\mathcal{F}$, and the metric (\ref{metric}), one finds the
following differential equation
\begin{equation}
F_{\varphi \rho }+\rho F_{\varphi \rho }^{\prime }=0,  \label{Fpr}
\end{equation}%
where $F_{\varphi \rho }=h^{\prime }(\rho )$ in which the prime denotes
differentiation with respect to $\rho $. Equation (\ref{Fpr}) has the
following solution
\begin{equation}
F_{\varphi \rho }=\frac{q}{\rho },  \label{Linfield}
\end{equation}%
where $q$ is an integration constant. To find the metric function $g(\rho )$%
, one may insert Eq. (\ref{Linfield}) in the field equation (\ref{Field
equation}) by considering the metric (\ref{metric}). After some
calculations, one can obtain the following differential equations
\begin{equation}
\left\{
\begin{array}{cc}
g^{\prime }(\rho )+2\Lambda \rho -\frac{2}{\rho }\left( \frac{q}{l}\right)
^{2}-cc_{1}m^{2}=0, & \rho \rho \ (\varphi \varphi )\ \;{component} \\
g^{\prime \prime }(\rho )+2\Lambda +2\left( \frac{q}{\rho l}\right) ^{2}=0,
& tt\ \;{component}%
\end{array}%
\right. ,
\end{equation}%
where the double prime is the second derivative versus $\rho $. It is
straightforward to show that these equations have the following solution
\begin{equation}
g(\rho )=m_{0}-\Lambda \rho ^{2}+\frac{2q^{2}}{l^{2}}\ln \left( \frac{\rho }{%
l}\right) +cc_{1}m^{2}\rho ,  \label{Linmetric}
\end{equation}%
which $m_{0}$ is an integration constant which is related to the
mass parameter, and $l$ is an arbitrary constant with length
dimension which is coming from the fact that the logarithmic
arguments should be dimensionless. As one can see, the massive
parameter appears in the metric function as a factor for the
linear function of $\rho $. We should note that the obtained
metric function (\ref{Linmetric}) satisfies all components of the
field equation (\ref{Field equation}), simultaneously. In
addition, the asymptotical behavior of the solution
(\ref{Linmetric}) is adS or dS provided $\Lambda <0$ or $\Lambda
>0$. Also, it is worthwhile to mention
that in the absence of massive parameter ($m=0$), the metric function (\ref%
{Linmetric}) reduces to the result of Ref. \cite{ThreeDim} for $s=1$.

\begin{center}
\textbf{A: Energy Conditions}
\end{center}

Now, we examine the energy conditions to find physical solutions. To do so,
we consider the orthonormal contravariant basis vectors, and then we obtain
the three dimensional energy momentum tensor as $T^{\mu \nu }=diag(\mu
,p_{r},p_{t})$ in which $\mu $, $p_{r}$, and $p_{t}$ are the energy density,
the radial pressure and the tangential pressure, respectively. Having the
energy momentum tensor at hand, we are in a position to investigate the
energy conditions. We use the following known constraints in three dimensions

\begin{center}
\begin{tabular}{cc}
\hline\hline
$%
\begin{array}{c}
p_{r}+\mu \geq 0 \\
p_{t}+\mu \geq 0%
\end{array}%
,$ & $\;{for null\; energy\; condition\; (NEC)}$ \vspace{0.2cm} \\ \hline
$%
\begin{array}{c}
\mu \geq 0 \\
p_{r}+\mu \geq 0 \\
p_{t}+\mu \geq 0%
\end{array}%
,$ & $\;{for\; weak \;energy\; condition\; (WEC)}$ \vspace{0.2cm} \\ \hline
$%
\begin{array}{c}
\mu \geq 0 \\
-\mu \leq p_{r}\leq \mu \\
-\mu \leq p_{t}\leq \mu%
\end{array}%
,$ & $\;{for\; dominant\; energy\; condition\; (DEC)}$ \vspace{0.2cm} \\
\hline
$%
\begin{array}{c}
p_{r}+\mu \geq 0 \\
p_{t}+\mu \geq 0 \\
\mu +p_{r}+p_{t}\geq 0%
\end{array}%
,$ & $\;{for\; strong \;energy\; condition\; (SEC)}$ \\ \hline\hline
\end{tabular}
\\[0pt]
\vspace{0.1cm} Table ($1$): Energy conditions criteria \vspace{0.5cm}
\end{center}

In order to simplify the mathematics and physical interpretations, we use
the following orthonormal contravariant (hatted) basis vectors for diagonal
static metric (\ref{metric})
\begin{equation}
\mathbf{e}_{\widehat{t}}=\frac{l}{\rho }\frac{\partial }{\partial t},\;{\ \
\ }\mathbf{e}_{\widehat{\rho }}=\sqrt{g}\frac{\partial }{\partial \rho },\ \
\mathbf{e}_{\widehat{\phi }}=\frac{1}{l\sqrt{g}}\frac{\partial }{\partial
\phi }.  \label{basis}
\end{equation}

It is a matter of straightforward calculations to show that the nonzero
components of stress-energy tensor are
\begin{equation}
T^{_{\widehat{t}\widehat{t}}}=T^{_{\widehat{\rho }\widehat{\rho }}}=T^{_{%
\widehat{\phi }\widehat{\phi }}}=\left( \frac{F_{\phi \rho }}{l}\right) ^{2}.
\label{EnergyCon}
\end{equation}

\begin{figure}[tbp]
$%
\begin{array}{c}
\resizebox{0.4\textwidth}{!}{ \includegraphics{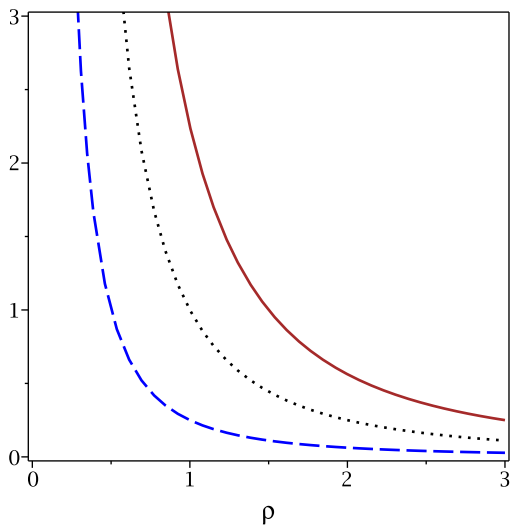}}%
\end{array}
$%
\caption{$T^{_{\widehat{t}\widehat{t}}}$ versus $\protect\rho$ for $l=1$ and
$q=0.5$ (dashed line), $q=1.0$ (doted line) and $q=1.5$ (continuous line).}
\label{Fig01}
\end{figure}

All components of stress-energy tensor are the same and positive and it is
easy to find that NEC, WEC, DEC and SEC are satisfied, simultaneously.

As one can see, the massive parameter do not contribute to the
energy-momentum tensor, so the energy conditions are independent of the
massive parameter. In order to investigation the effects of charge on the
energy density of the spacetime, we plot the $T^{\widehat{t}\widehat{t}}$\
versus $\rho $. Considering Fig. \ref{Fig01}, one can find that the energy
density of the spacetime is positive everywhere, and increasing the charge
parameter leads to increasing the concentration of energy density.

\begin{center}
\textbf{B: Geometric Properties}
\end{center}

Now, we want to study the properties of spacetime described by Eq. (\ref%
{metric}) with obtained metric function (\ref{Linmetric}). At first, we
calculate $R_{\mu \nu \lambda \kappa }R^{\mu \nu \lambda \kappa }$ for
examination of existence of curvature singularity
\begin{equation}
R_{\mu \nu \lambda \kappa }R^{\mu \nu \lambda \kappa }=12\Lambda ^{2}-\frac{%
8\Lambda q^{2}}{l^{2}\rho ^{2}}+\frac{2cc_{1}m^{2}(4q^{2}+cc_{1}m^{2}l^{2}%
\rho -4\Lambda l^{2}\rho ^{2})}{l^{2}\rho ^{3}}+\frac{12q^{4}}{l^{4}\rho ^{4}%
}.  \label{LinKretschmann}
\end{equation}

\begin{figure}[tbp]
\resizebox{0.8\textwidth}{!}{  \includegraphics{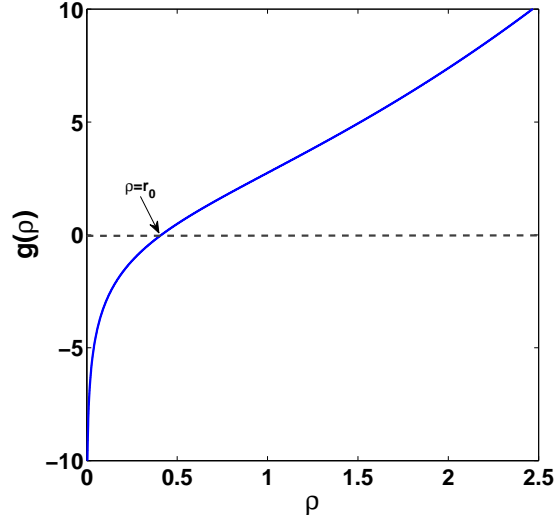}}
\caption{$g(\protect\rho )$ versus $\protect\rho $ for $l=1$, $q=1$, $%
\Lambda =-1$, $c=1$, $c_{1}=1$, $m=0.5$ and $m_{0}=1.5$.}
\label{Fig011}
\end{figure}

Considering Eq. (\ref{LinKretschmann}), the Kretschmann scalar reduces to $%
12\Lambda ^{2}$\ for $\rho \longrightarrow \infty $, which confirms that the
asymptotical behavior of this spacetime is (a)dS. It is also obvious that
the Kretschmann scalar diverges at $\rho =0$, and therefore one might think
that there is a curvature singularity located at $\rho =0$. But as we will
see, the spacetime will never achieve $\rho =0$. There are two possible
cases for the metric function: the metric function has no real positive root
which is interpreted as naked singularity (this case is not of interest
here), or metric function has at least one real positive root. If one
considers $r_{0}$\ as the largest root of metric function, it is clear that
for $\rho <r_{0}$\ there will be a change in signature of metric (see Fig. %
\ref{Fig011}). In other words, for $\rho <r_{0}$\ the metric function is
negative, hence metric signature is $(-,-,-)$, and for $\rho >r_{0}$\ the
metric function is positive, therefore metric signature is legal $(-,+,+)$.
This change in the metric signature results into a conclusion: it is not
possible to extend spacetime to $\rho <r_{0}$. In order to exclude the
forbidden zone ($\rho <r_{0}$), we introduce a new radial coordinate $r$ as
\begin{equation}
r^{2}=\rho ^{2}-r_{0}^{2}\Longrightarrow d\rho ^{2}=\frac{r^{2}}{%
r^{2}+r_{0}^{2}}dr^{2},  \label{coordinate}
\end{equation}%
where for the allowed region, $\rho \geq r_{0}$, leads to $r\geq 0$ in the
new coordinate system. Applying this coordinate transformation, the metric (%
\ref{metric}) should be written as
\begin{equation}
ds^{2}=-\frac{r^{2}+r_{0}^{2}}{l^{2}}dt^{2}+\frac{r^{2}}{\left(
r^{2}+r_{0}^{2}\right) g(r)}dr^{2}+l^{2}g(r)d\varphi ^{2},
\label{change coordinate metric}
\end{equation}%
%
%
%
%
%
%
%
%
%
%
in which the coordinate $r$ assumes the values $0\leq r<\infty $, and
obtained $g(r)$ (Eq. (\ref{Linmetric})) is now given by
\begin{equation}
g(r)=m_{0}-\Lambda \left( r^{2}+r_{0}^{2}\right) +\frac{q^{2}}{l^{2}}\ln
\left( \frac{r^{2}+r_{0}^{2}}{l^{2}}\right) +cc_{1}m^{2}\sqrt{r^{2}+r_{0}^{2}%
}.  \label{change Linmetric}
\end{equation}

The nonzero component of electromagnetic field in the new coordinate can be
given by
\begin{equation}
F_{\varphi r}=\frac{q}{\sqrt{r^{2}+r_{0}^{2}}},  \label{new Linfield}
\end{equation}

One can show that all curvature invariants do not diverge in the range $%
0\leq r<\infty $, and $g(r)$ (Eq. (\ref{change Linmetric})) is positive
definite for $0\leq r<\infty $. It is evident that for having singular
solutions both $r$ and $r_{0}$ must be zero whereas this case is never
reached due to considering nonzero value for $r_{0}.$ So, this spacetime has
no curvature singularity and horizon. Due to the fact that the limit of the
ratio "circumference/radius" is not $2\pi $, the spacetime (\ref{change
coordinate metric}) has a conic geometry and therefore the spacetime has a
conical singularity at $r=0$
\begin{equation}
\lim_{r\longrightarrow 0}\frac{1}{r}\sqrt{\frac{g_{\varphi \varphi }}{g_{rr}}%
}\neq 1.
\end{equation}

On the other hand, the conical singularity can be removed if one exchanges
the coordinate $\varphi $ with the following period
\begin{equation}
Period_{\varphi }=2\pi \left( \lim_{r\longrightarrow 0}\frac{1}{r}\sqrt{%
\frac{g_{\varphi \varphi }}{g_{rr}}}\right) ^{-1}=2\pi \left( 1-4\mu \right)
,  \label{Period}
\end{equation}%
in which the deficit angle is defined as $\delta \varphi =8\pi \mu
$, where $\mu $ is given by
\begin{equation}
\mu =\frac{1}{4}+\frac{1}{4lr_{0}\Omega },  \label{miu}
\end{equation}%
where $\Omega $ is
\begin{equation}
\Omega =\Lambda -\frac{q^{2}}{l^{2}r_{0}^{2}}-\frac{cc_{1}m^{2}}{2r_{0}}.
\label{omega}
\end{equation}

In order to have a better insight of the behavior of deficit angle, we
calculate the root and divergence points of the deficit angle as
\begin{equation}
\left. r_{0}\right\vert _{\delta \varphi =0}=\left\{
\begin{array}{c}
\frac{1}{4l\Lambda }\left( cc_{1}m^{2}l-2\pm \sqrt{%
c^{2}c_{1}^{2}m^{4}l^{2}-4\left( cc_{1}m^{2}l-1-4\Lambda q^{2}\right) }%
\right) \\
\\
\frac{1}{4l\Lambda }\left( -cc_{1}m^{2}l-2\pm \sqrt{%
c^{2}c_{1}^{2}m^{4}l^{2}-4\left( cc_{1}m^{2}l-1-4\Lambda q^{2}\right) }%
\right)%
\end{array}%
\right. ,  \label{rootMax}
\end{equation}

\begin{equation}
\left. r_{0}\right\vert _{\delta \varphi \longrightarrow \infty }=\left\{
\begin{array}{c}
\frac{1}{4l\Lambda }\left( cc_{1}m^{2}l\pm \sqrt{c^{2}c_{1}^{2}m^{4}l^{2}+16%
\Lambda q^{2}}\right) \\
\\
\frac{1}{4l\Lambda }\left( -cc_{1}m^{2}l\pm \sqrt{%
c^{2}c_{1}^{2}m^{4}l^{2}+16\Lambda q^{2}}\right)%
\end{array}%
\right. .  \label{divMax}
\end{equation}

Here, we see that the roots are functions of the cosmological constant,
massive gravity and electric charge. Existence of the real valued root is
restricted to following condition
\begin{equation}
c^{2}c_{1}^{2}m^{4}l^{2}-4\left( cc_{1}m^{2}l-1-4\Lambda q^{2}\right) \geq 0.
\end{equation}

The effects of the massive gravity and electric charge are only observed in
numerator of the roots while the effects of the cosmological constant could
be observed in both numerator and denominator of the roots. The electric
charge is coupled with cosmological constant. While such coupling is not
observed for the massive gravity.

As for the divergencies of the deficit angle, one can observe that its
existence is also restricted to satisfaction of specific condition in the
following form
\begin{equation}
c^{2}c_{1}^{2}m^{4}l^{2}+16\Lambda q^{2}\geq 0
\end{equation}

In the absence of the massive gravity, only for dS spacetime divergencies
are observable for deficit angle. Generalization to massive gravity provides
the possibility of the divergencies for deficit angle in adS spacetime under
certain circumstances. This highlights the effects of the massive gravity.
Here, similar to the case of roots, a coupling between cosmological constant
and electric charge is observed while such coupling could not be seen for
massive gravity.

\section{Generalization of achievements to the case of nonlinear
electrodynamics: PMI theory}

In this section, we are going to obtain the solutions in presence of PMI
source and investigate the properties. We start with the following PMI
Lagrangian
\begin{equation}
\mathcal{L}_{PMI}(\mathcal{F})=(-\kappa \mathcal{F})^{s},  \label{PMI}
\end{equation}%
where $\kappa $ and $s$ are coupling and power constants, respectively.
Obviously, the PMI Lagrangian (\ref{PMI}) reduces to the standard Maxwell
Lagrangian ($\mathcal{L}_{Maxwell}(\mathcal{F})=-\mathcal{F}$) for $s=1$ and
$\kappa =1$ which we have investigated before.

Following the method of previous section and considering Eqs.
(\ref{Maxwell equation}), (\ref{metric}) and (\ref{PMI}), one can
obtain the following differential equation for nonzero component
of Faraday tensor
\begin{equation}
F_{\varphi \rho }+(2s-1)\rho F_{\varphi \rho }^{\prime }=0,  \label{Diff}
\end{equation}%
with the following solution%
\begin{equation}
F_{\varphi \rho }=q\rho ^{1/(1-2s)},  \label{Lagrangian field}
\end{equation}%
in which $q$ is an integration constant. In order to have a physical
asymptotical behavior, we should consider $s>1/2$.\ On the other hand, one
can easily show that the vector potential $A_{\varphi }$, is%
\begin{equation}
A_{\varphi }=q\rho ^{\frac{2\left( s-1\right) }{2s-1}},
\end{equation}%
the electromagnetic gauge potential should be finite at infinity ($\rho
\rightarrow \infty $), therefore, one should impose following restriction to
have this property, so we have%
\begin{equation}
\frac{2\left( s-1\right) }{2s-1}<0.
\end{equation}

The above equation leads to the following restriction on the range of $s$, as%
\begin{equation}
\frac{1}{2}<s<1.
\end{equation}

Here, one can insert Eq. (\ref{Lagrangian field}) in the gravitational field
equation (\ref{Field equation}) by considering the metric (\ref{metric}) to
obtain the metric function $g(\rho )$ as
\begin{equation}
g(\rho )=m_{0}-\Lambda \rho ^{2}+cc_{1}m^{2}\rho +\frac{(2s-1)^{2}\rho ^{2}}{%
2(s-1)}\chi (\rho ),  \label{metricfun}
\end{equation}%
where%
\begin{equation}
\chi (\rho )=\left( \frac{2q^{2}}{l^{2}}\rho ^{2/(1-2s)}\right) ^{s}.
\end{equation}

It is notable that, the obtained metric function in Eq. (\ref{metricfun}) is
related to $s\neq 1$. Also, $m_{0}$ is an integration constant which is
related to the mass of solutions.

Now, one can calculate the nonzero components of stress-energy tensor by
using the introduced basis vectors in Eq. (\ref{basis}) as
\begin{eqnarray}
T^{_{\widehat{t}\widehat{t}}} &=&\frac{1}{2}\left( \frac{2F_{\phi \rho }^{2}%
}{l^{2}}\right) ^{s},  \label{EnergyConPMI} \\
&&  \notag \\
T^{_{\widehat{\rho }\widehat{\rho }}} &=&T^{_{\widehat{\phi }\widehat{\phi }%
}}=\left( s-\frac{1}{2}\right) \left( \frac{2F_{\phi \rho }^{2}}{l^{2}}%
\right) ^{s}.
\end{eqnarray}

\begin{figure}[tbp]
$%
\begin{array}{cc}
\resizebox{0.35\textwidth}{!}{ \includegraphics{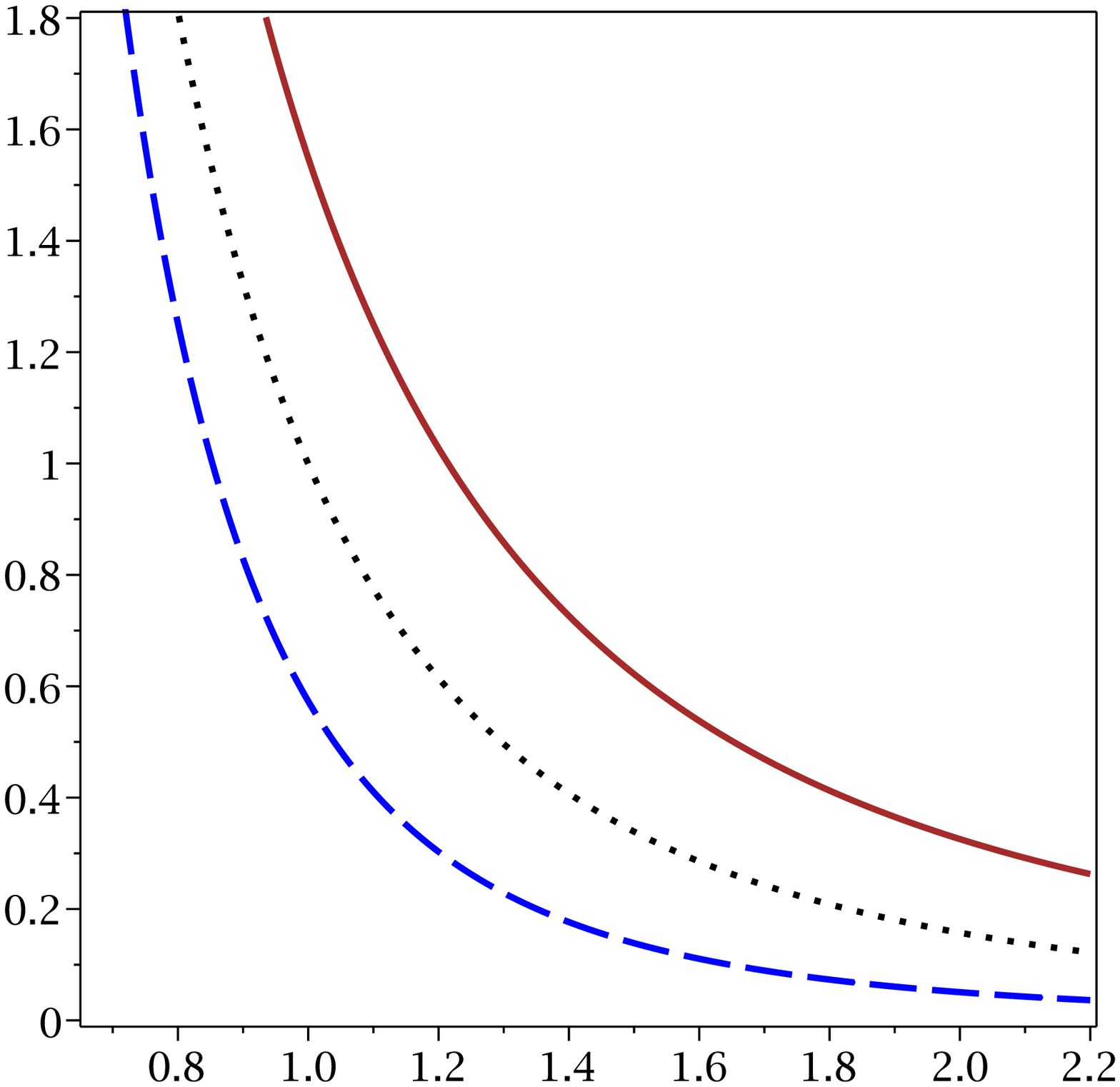}} & %
\resizebox{0.35\textwidth}{!}{ \includegraphics{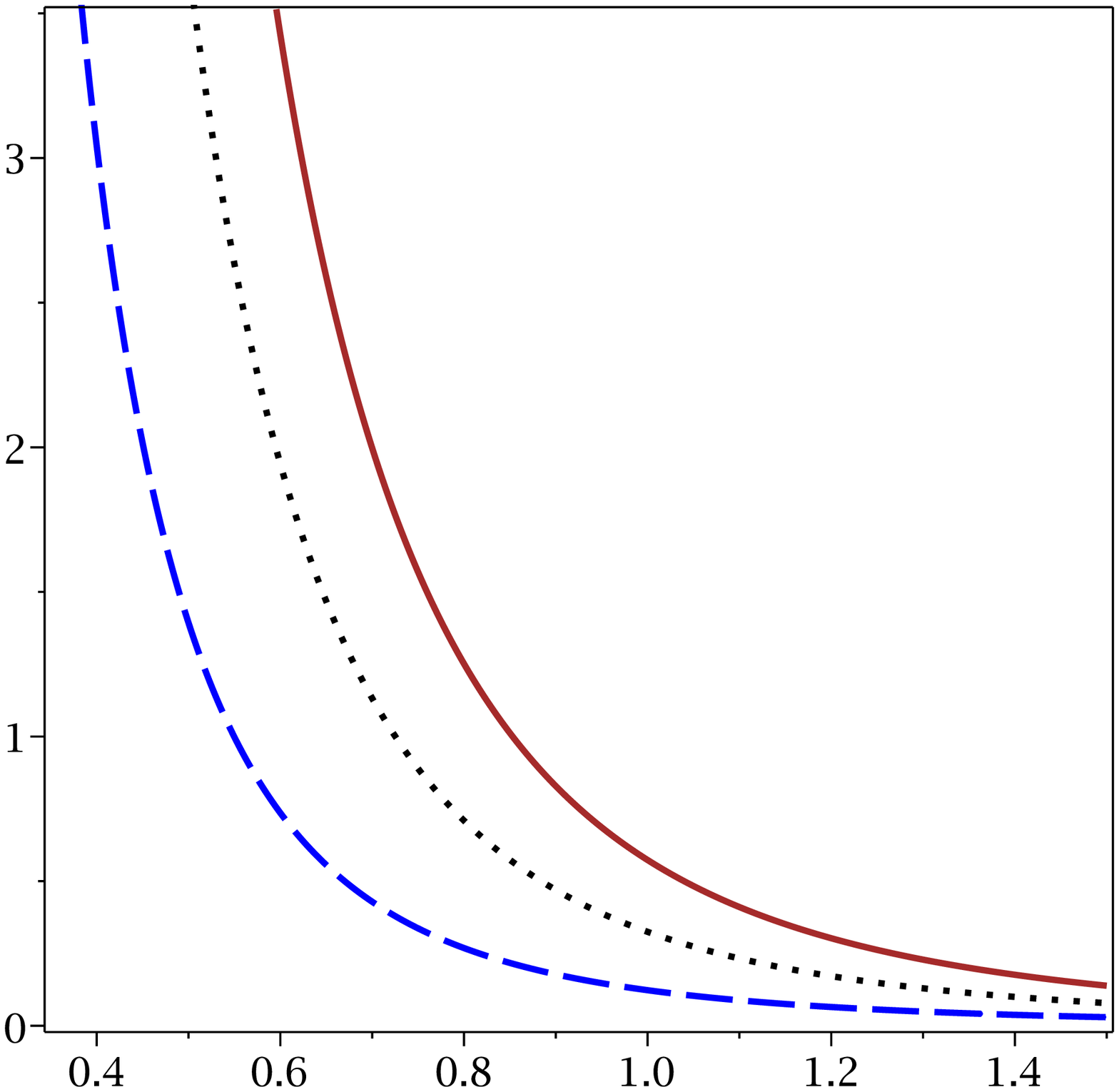}}%
\end{array}
$%
\caption{$T^{_{\widehat{t}\widehat{t}}}$ versus $\protect\rho $ for $l=1$.
\newline
\textbf{Left diagram:} for $q=1.5$, $s=0.7$ (dashed line), $s=0.8$ (doted
line) and $s=0.9$ (continuous line). \newline
\textbf{Right diagram:} for $s=0.7$, $q=0.5$ (dashed line), $q=1.0$ (doted
line) and $q=1.5$ (continuous line).}
\label{Fig02}
\end{figure}

According to the above equation, $\mu $ ($T^{_{\widehat{t}\widehat{t}}}$) is
positive, and so the NEC, WEC, and SEC are satisfied, simultaneously. In
addition, in order to satisfy the DEC, the parameter of PMI ($s$) must be in
the range $\frac{1}{2}<s<1$. As we have mentioned before, the energy
conditions do not depend on the massive parameter. Here, we want to
investigate the effects of PMI parameter ($s$) and electrical charge ($q$)
on the energy conditions, so we plot $T^{_{\widehat{t}\widehat{t}}}$ versus $%
\rho $ in Fig. \ref{Fig02}. As one can see, increasing the parameter of PMI
theory and electrical charge leads to increasing the concentration of energy
density.

One can show that the metric (\ref{metric}) with the metric function (\ref%
{metricfun})\ has a singularity at $\rho =0$ by calculating the Kretschmann
scalar as
\begin{eqnarray}
R_{\mu \nu \lambda \kappa }R^{\mu \nu \lambda \kappa } &=&12\Lambda ^{2}-%
\frac{4cc_{1}m^{2}}{\rho }\left[ 2\Lambda -(2s-1)\chi _{q,\rho ,s}-\frac{%
cc_{1}m^{2}}{2\rho }\right]  \notag \\
&&  \notag \\
&&+\left[ (8s^{2}-8s+3)\chi _{q,\rho ,s}-4(4s-3)\Lambda \right] \chi
_{q,\rho ,s}.  \label{kre}
\end{eqnarray}

From Eq. (\ref{kre}), it is obvious that the Kretschmann scalar reduces to $%
12\Lambda ^{2}$\ for $\rho \longrightarrow \infty $\ and diverges at $\rho
=0 $. On the other hand, as we mentioned before, it is not possible to
extend spacetime to $\rho <r_{0}$ because of signature changing. Also, one
can apply the coordinate transformation (\ref{coordinate}) to the metric (%
\ref{metric}) and find the metric function as
\begin{equation}
g(r)=m_{0}-\Lambda \left( r^{2}+r_{0}^{2}\right) +cc_{1}m^{2}\left(
r^{2}+r_{0}^{2}\right) ^{1/2}+\frac{(2s-1)^{2}\left( r^{2}+r_{0}^{2}\right)
}{2(s-1)}\chi (r),  \label{change Lagrangian metric}
\end{equation}%
where%
\begin{equation}
\chi (r)=\left( \frac{2q^{2}}{l^{2}}\left( r^{2}+r_{0}^{2}\right)
^{1/(1-2s)}\right) ^{s},
\end{equation}%
and the electromagnetic field in the new coordinate is
\begin{equation}
F_{\varphi r}=q\left( r^{2}+r_{0}^{2}\right) ^{1/2(1-2s)}.
\end{equation}

Since all curvature invariants do not diverge in the range $0\leq r<\infty $%
, one finds that there is no essential singularity. But like the Maxwell
case, this spacetime has a conical singularity at $r=0$ with the deficit
angle $\delta \varphi =8\pi \mu $ where $\mu $ is given by Eq. (\ref{miu})
and $\Omega $ has the following form
\begin{equation}
\Omega =\Lambda -\frac{cc_{1}m^{2}}{2r_{0}}-\frac{(2s-1)}{2}\left( \frac{%
2q^{2}}{l^{2}}r_{0}^{2/(1-2s)}\right) ^{s}.  \label{omegaPMI}
\end{equation}

Due to complexity of obtained relation in Eq. (\ref{omegaPMI}), it is not
possible to calculate the root and divergence points of deficit angle
analytically, therefore, we study them in some graphs in next section.

\section{deficit angle diagrams}

\begin{figure}[tbp]
$%
\begin{array}{cc}
\resizebox{0.35\textwidth}{!}{ \includegraphics{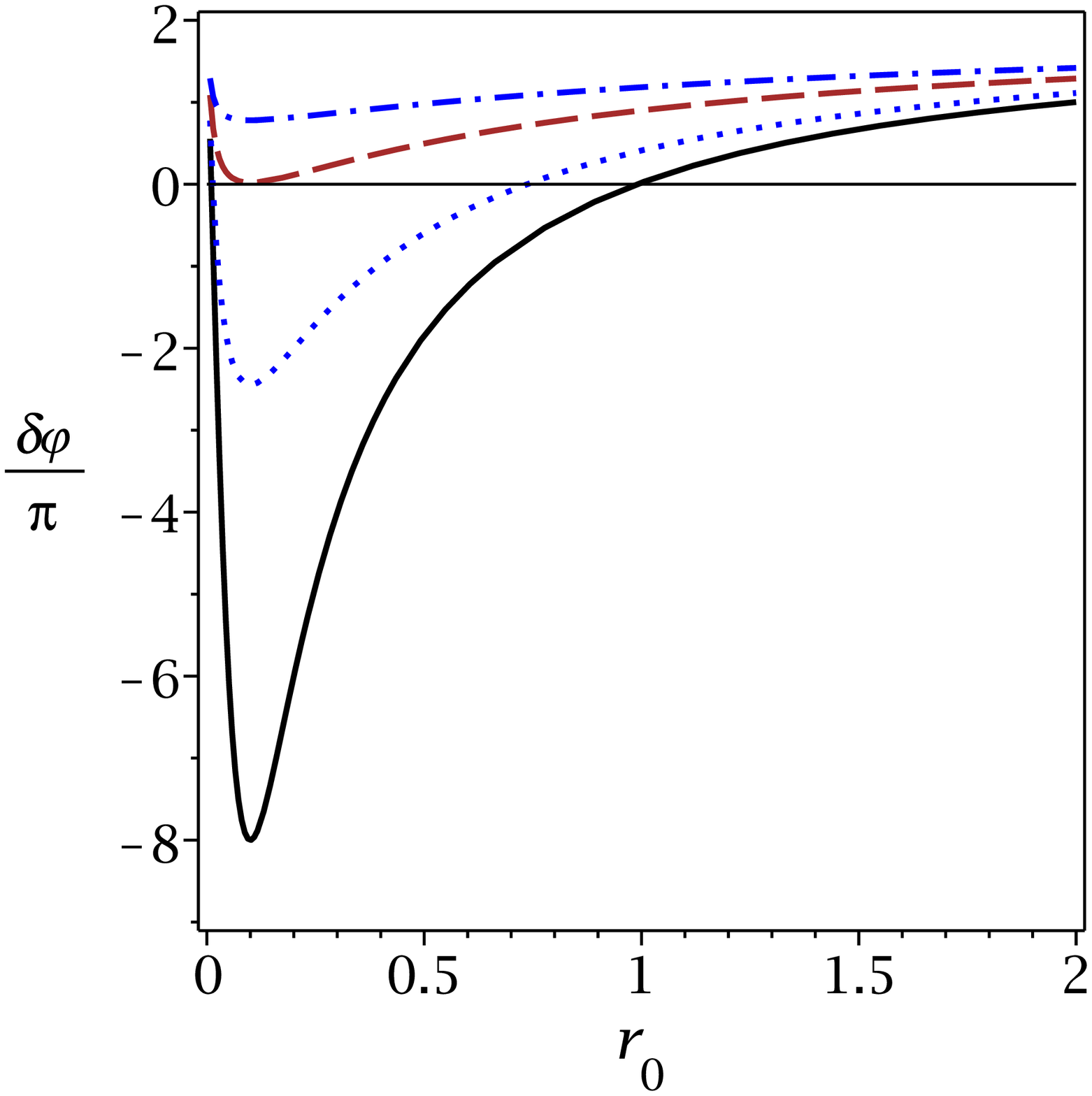}} & %
\resizebox{0.35\textwidth}{!}{ \includegraphics{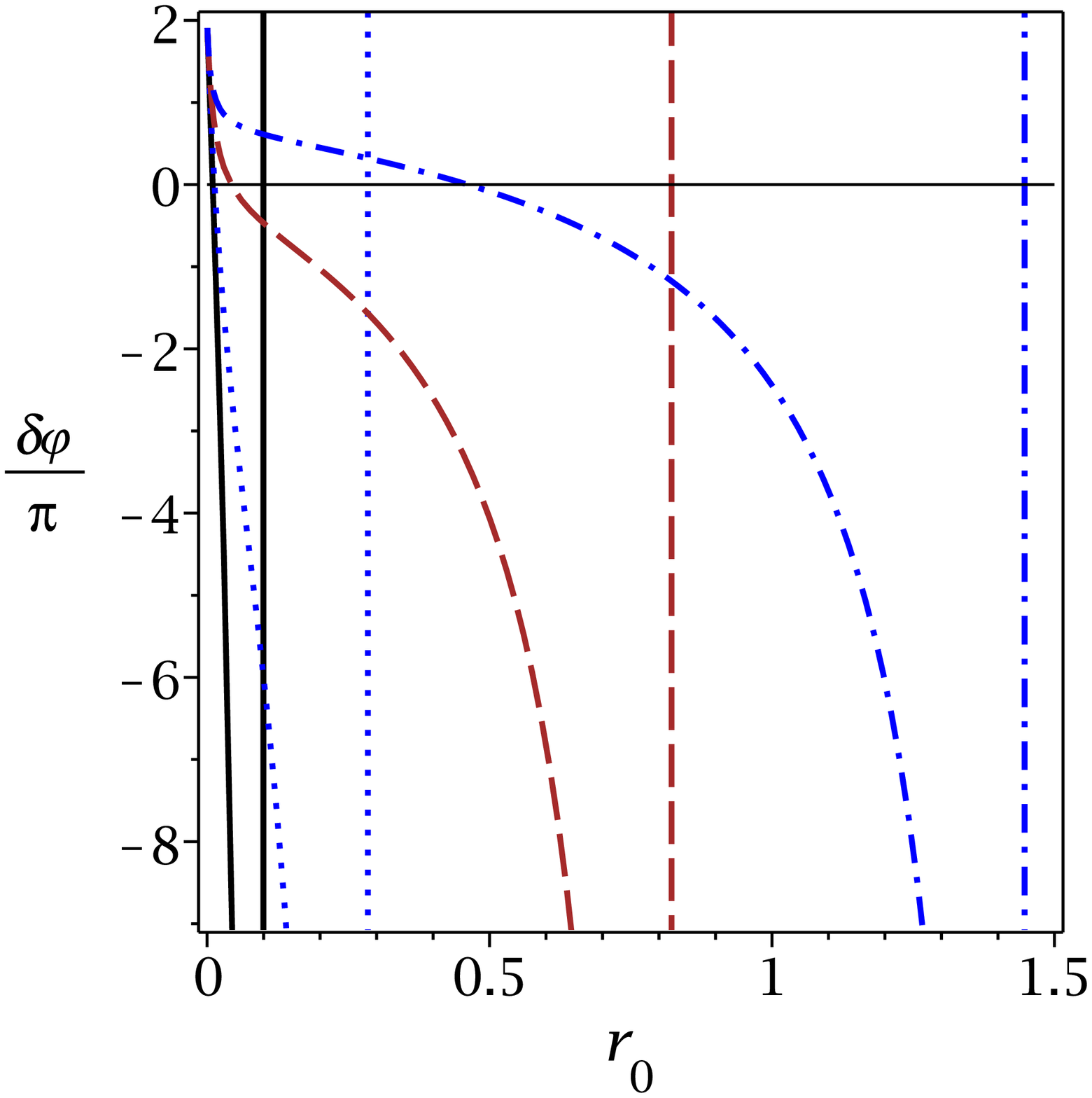}}%
\end{array}
$%
\caption{\textbf{\emph{Maxwell solutions:}} $\protect\delta \protect\varphi $
versus $r_{0}$ for $q=0.1$, $c=1$ and $c_{1}=2$, $m=0$ (continuous line), $%
m=0.5$ (dotted line), $m=0.9$ (dashed line) and $m=1.2$ (dashed-dotted
line). \newline
\textbf{Left diagram:} $\Lambda =-1$; \textbf{Right diagram:} $\Lambda =1$.}
\label{Fig4}
\end{figure}

\begin{figure}[tbp]
$%
\begin{array}{cc}
\resizebox{0.35\textwidth}{!}{ \includegraphics{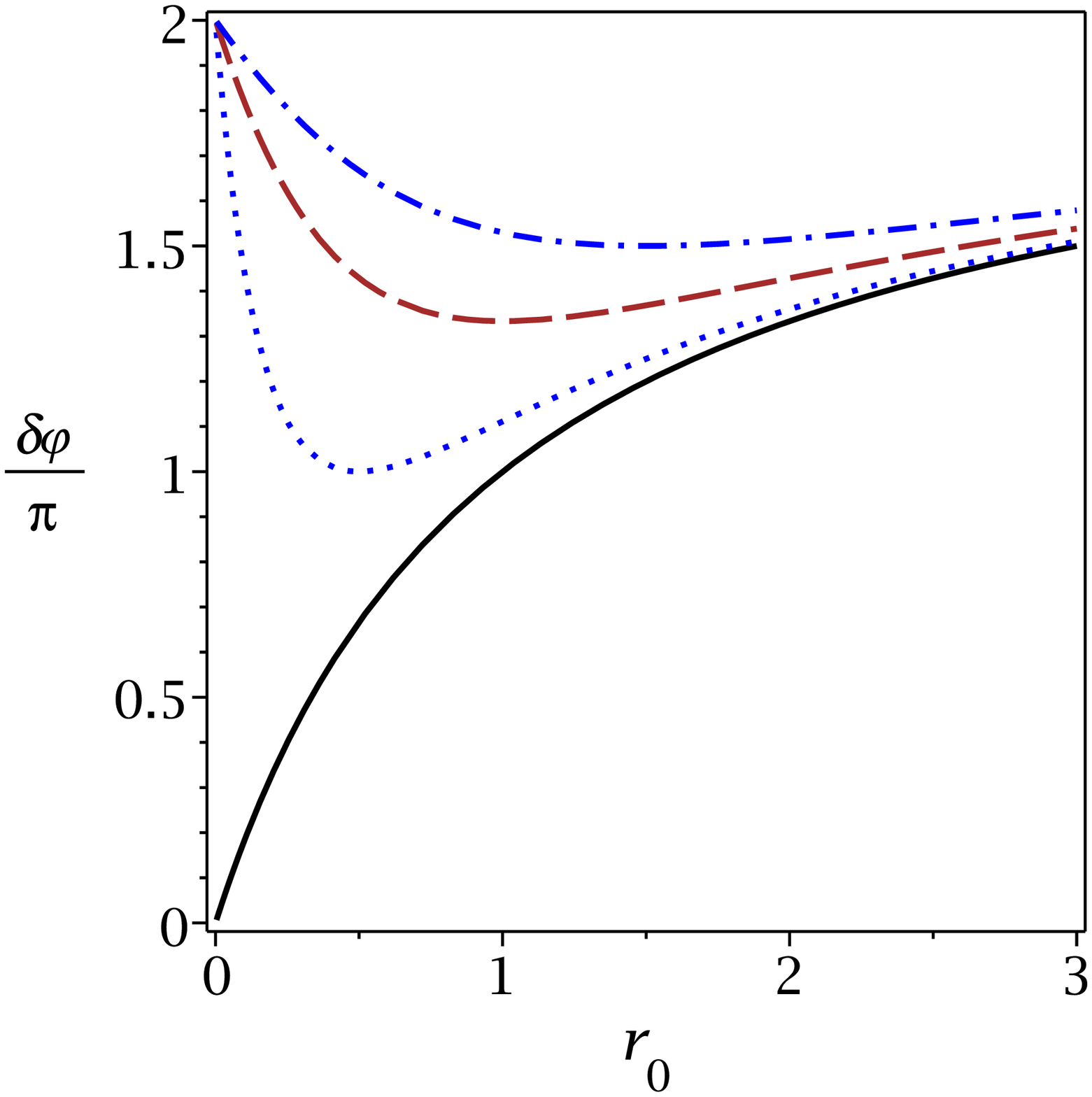}} & %
\resizebox{0.35\textwidth}{!}{ \includegraphics{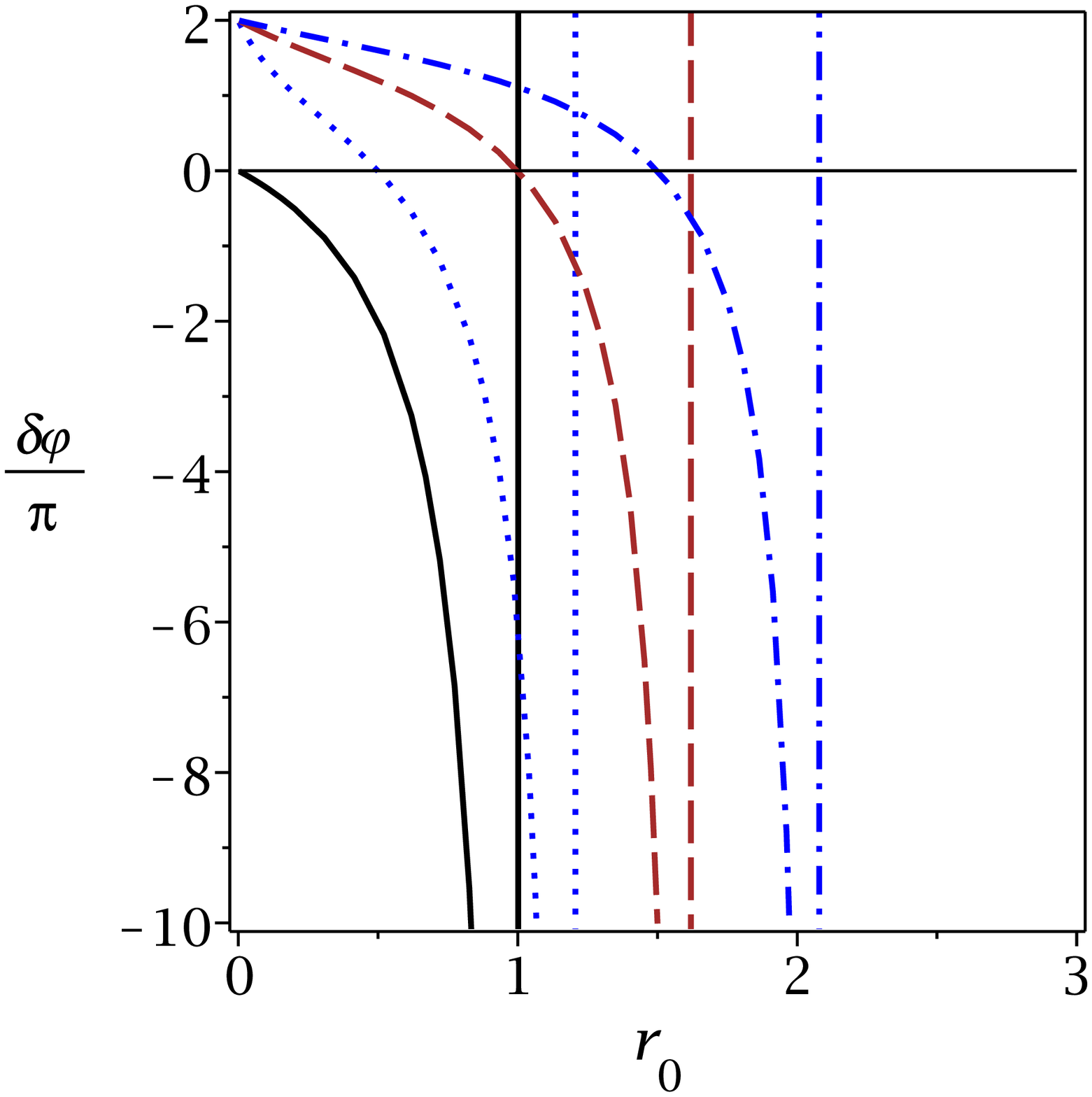}}%
\end{array}
$%
\caption{\textbf{\emph{Maxwell solutions:}} $\protect\delta \protect\varphi $
versus $r_{0}$ for $m=1$, $c=1$ and $c_{1}=2$, $q=0$ (continuous line), $%
q=0.5$ (dotted line), $q=1$ (dashed line) and $q=1.5$ (dashed-dotted line).
\newline
\textbf{Left diagram:} $\Lambda =-1$; \textbf{Right diagram:} $\Lambda =1$.}
\label{Fig5}
\end{figure}

\begin{figure}[tbp]
$%
\begin{array}{cc}
\resizebox{0.35\textwidth}{!}{ \includegraphics{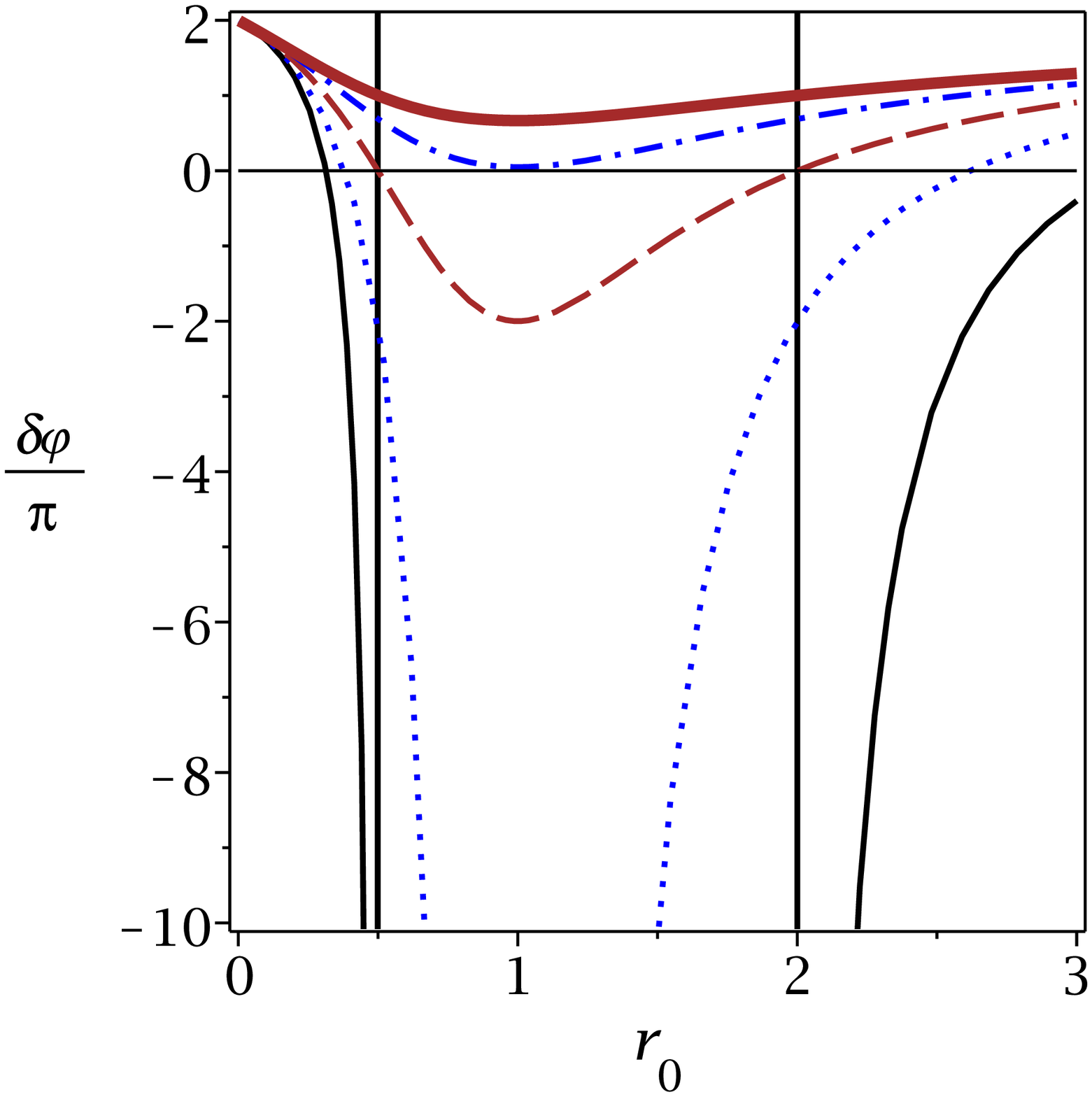}} & %
\resizebox{0.35\textwidth}{!}{ \includegraphics{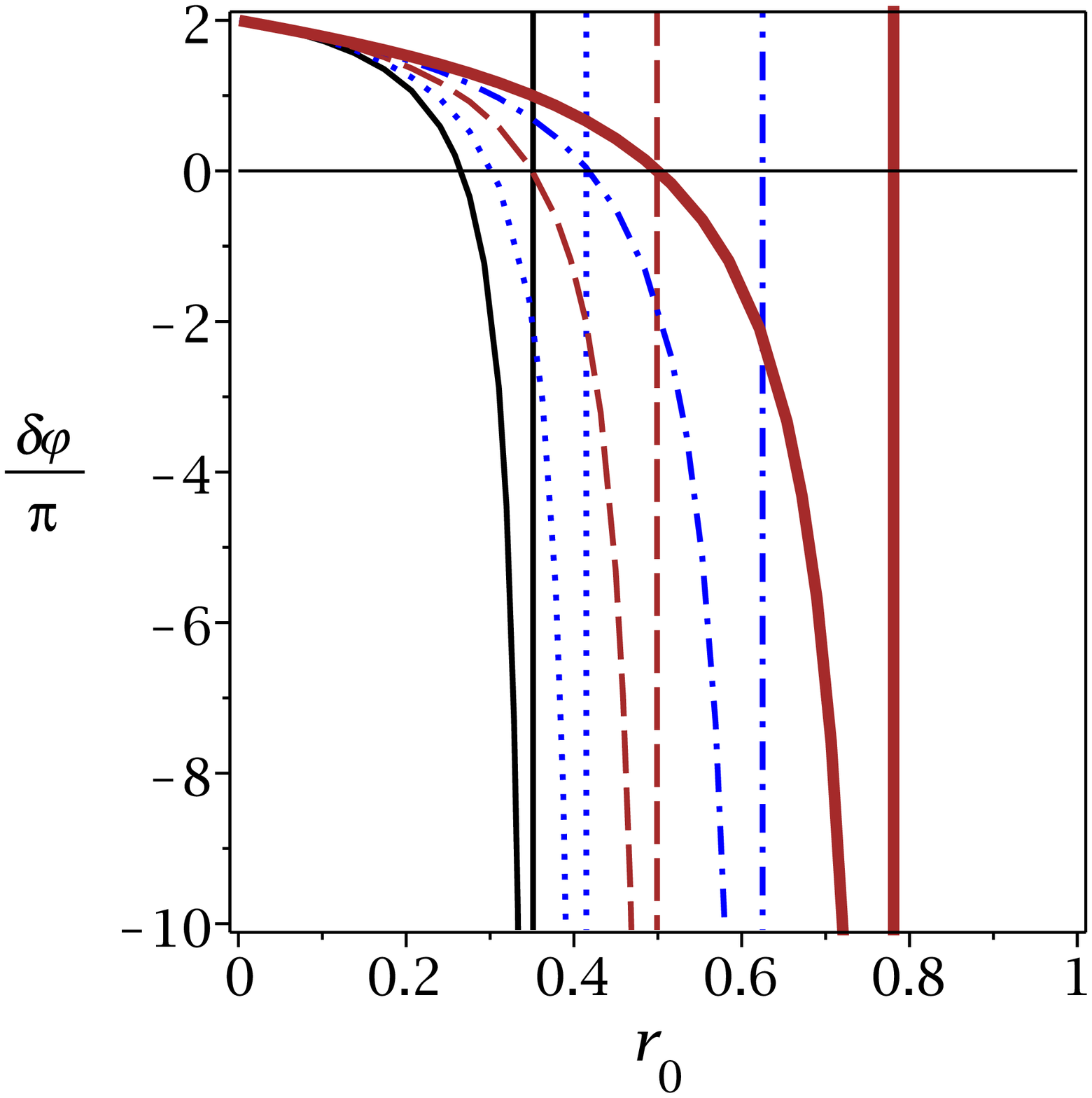}}%
\end{array}
$%
\caption{\textbf{\emph{Maxwell solutions:}} $\protect\delta \protect\varphi $
versus $r_{0}$ for $q=0.1$, $c=1$ and $m=1$, $c_{1}=-5$ (continuous line), $%
c_{1}=-4$ (dotted line), $c_{1}=-3$ (dashed line), $c_{1}=-1.95$
(dashed-dotted line) and $c_{1}=-1$ (bold line). \newline
\textbf{Left diagram:} $\Lambda =-1$; \textbf{Right diagram:} $\Lambda =1$.}
\label{Fig6}
\end{figure}

\begin{figure}[tbp]
$%
\begin{array}{cc}
\resizebox{0.35\textwidth}{!}{ \includegraphics{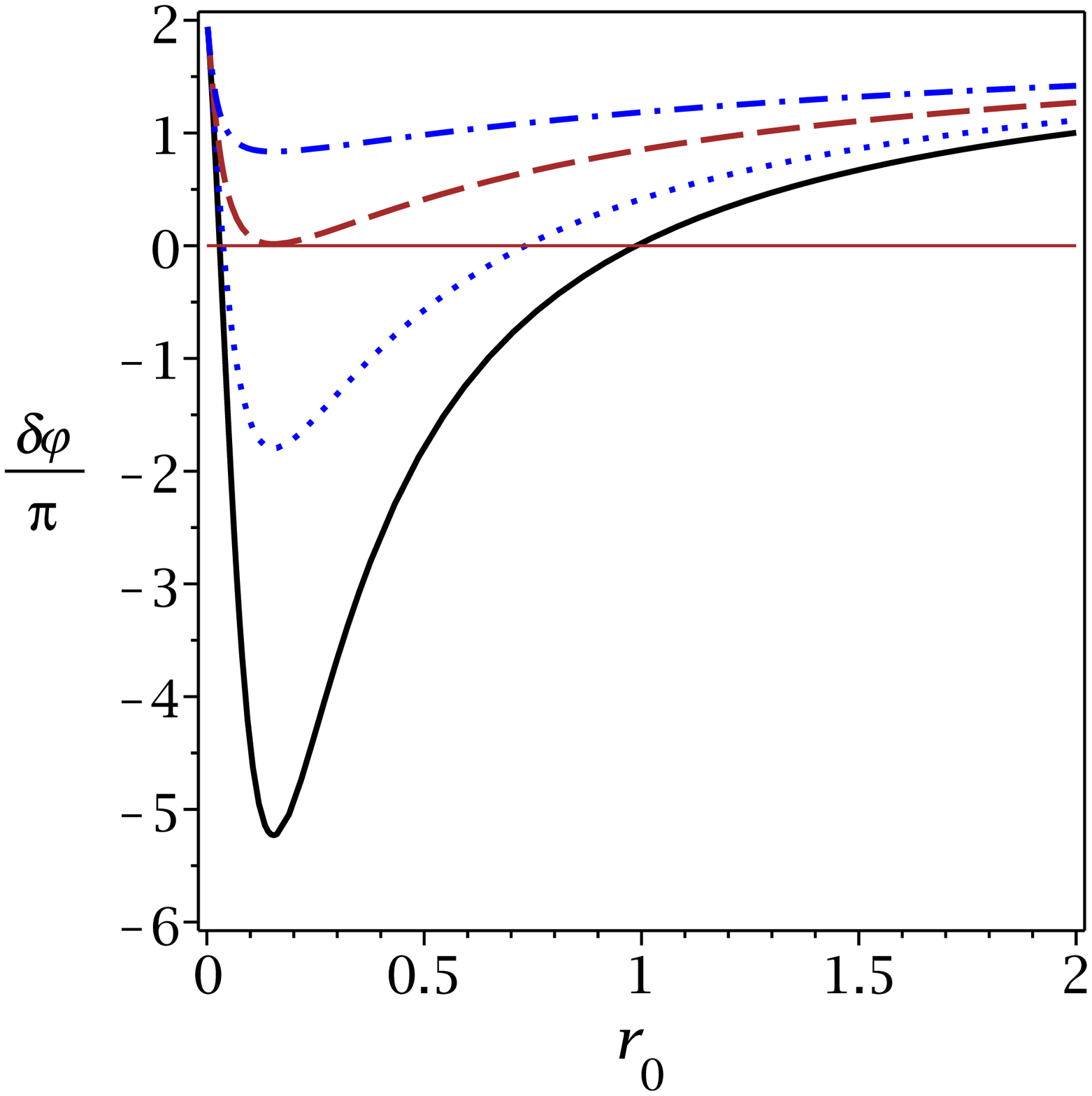}} & %
\resizebox{0.35\textwidth}{!}{ \includegraphics{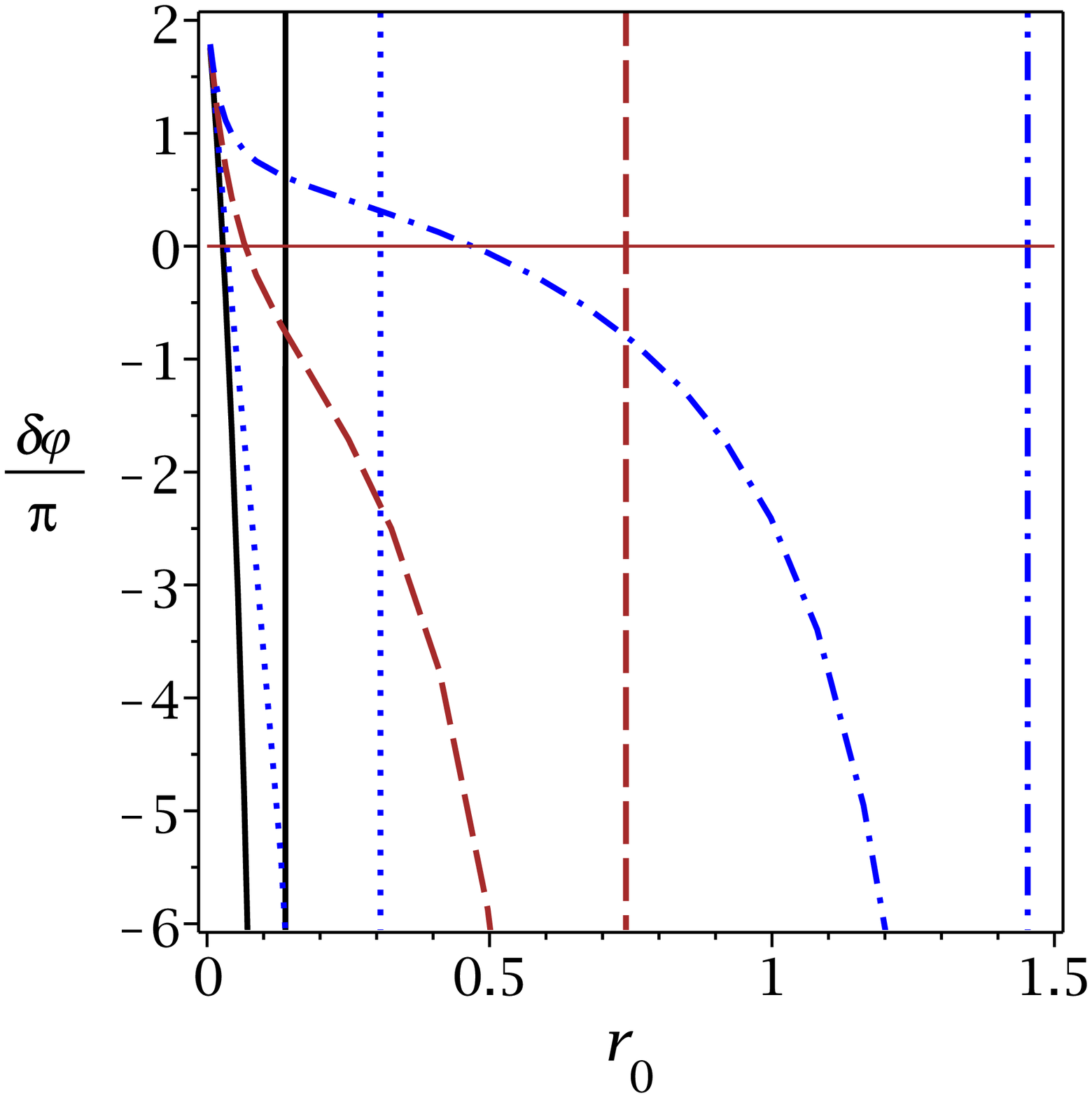}}%
\end{array}
$%
\caption{\textbf{\emph{PMI solutions:}} $\protect\delta \protect\varphi $
versus $r_{0}$ for $q=0.1$, $c=1$, $c_{1}=2$ and $s=0.9$, $m=0$ (continuous
line), $m=0.5$ (dotted line), $m=0.85$ (dashed line) and $m=1.2$
(dashed-dotted line). \newline
\textbf{Left diagram:} $\Lambda =-1$; \textbf{Right diagram:} $\Lambda =1$.}
\label{Fig7}
\end{figure}

In order to study the effects of different parameters on the properties of
deficit angle for the Maxwell and PMI cases, we have plotted various
diagrams (Figs. \ref{Fig4}-\ref{Fig6} for Maxwell case and Figs. \ref{Fig7}-%
\ref{Fig10} for PMI case). The left panels are dedicated to adS spacetime
while the right ones are related to dS spacetime. In Ref. \cite{Mamasani},
it was pointed out that in order to remove ensemble dependency, $l$ should
be replaced by following relation
\begin{equation}
\Lambda =\pm \frac{1}{l}, \label{Lam}
\end{equation}%
where the positive branch is related to dS spacetime and the opposite is for
adS solutions. Hereafter, we employ Eq. (\ref{Lam}) to plot deficit angle
diagrams. It is notable to highlight a few remarks regarding to values of
deficit angle. The deficit angle is restricted by an upper limit provided by
geometrical properties of the solutions. Its value could not exceed $2\pi$,
and more precisely, deficit angle could have values in range of $-\infty
<\delta \varphi \leq 2\pi$.

\begin{figure}[tbp]
$%
\begin{array}{cc}
\resizebox{0.35\textwidth}{!}{ \includegraphics{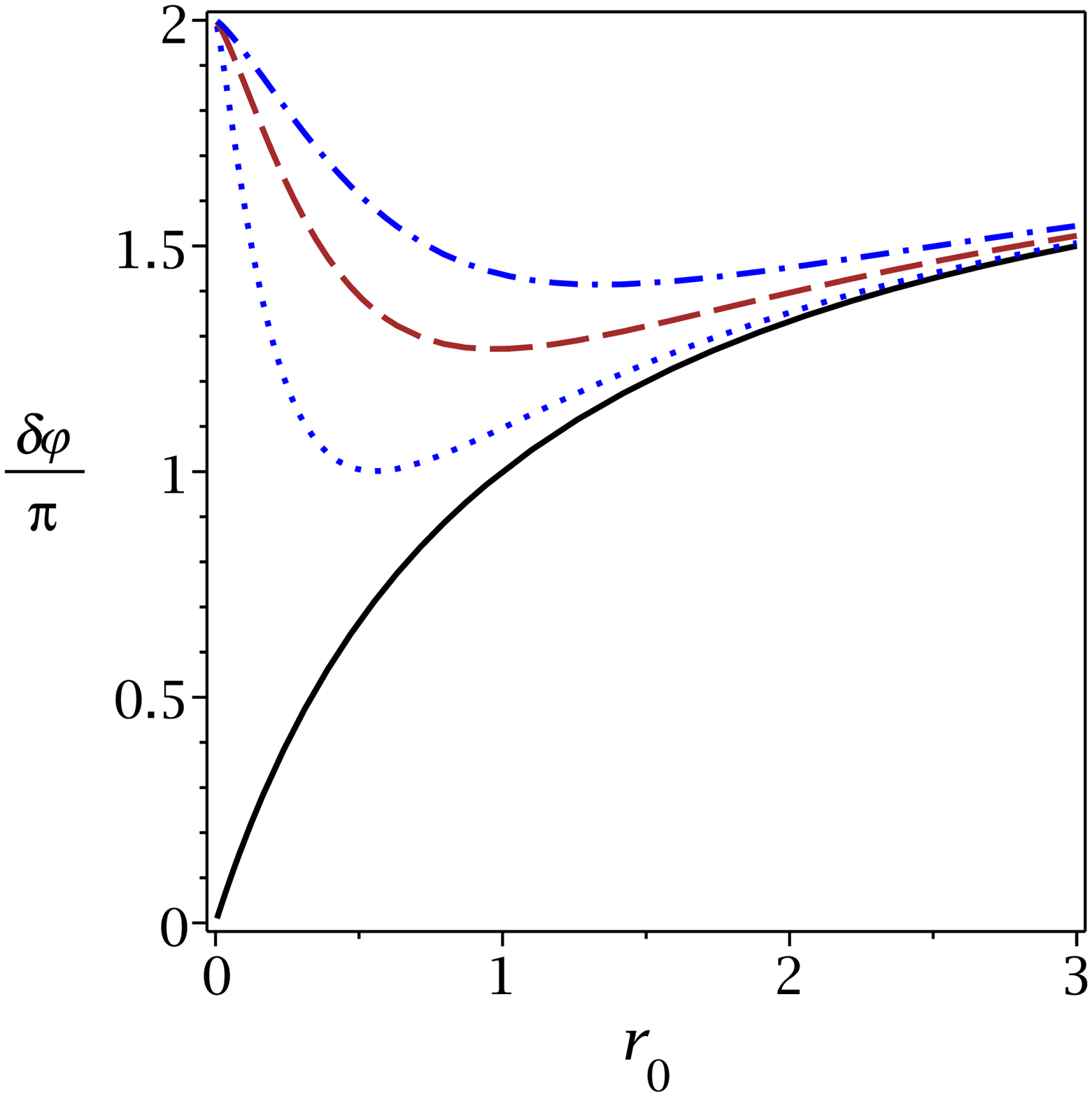}} & %
\resizebox{0.35\textwidth}{!}{ \includegraphics{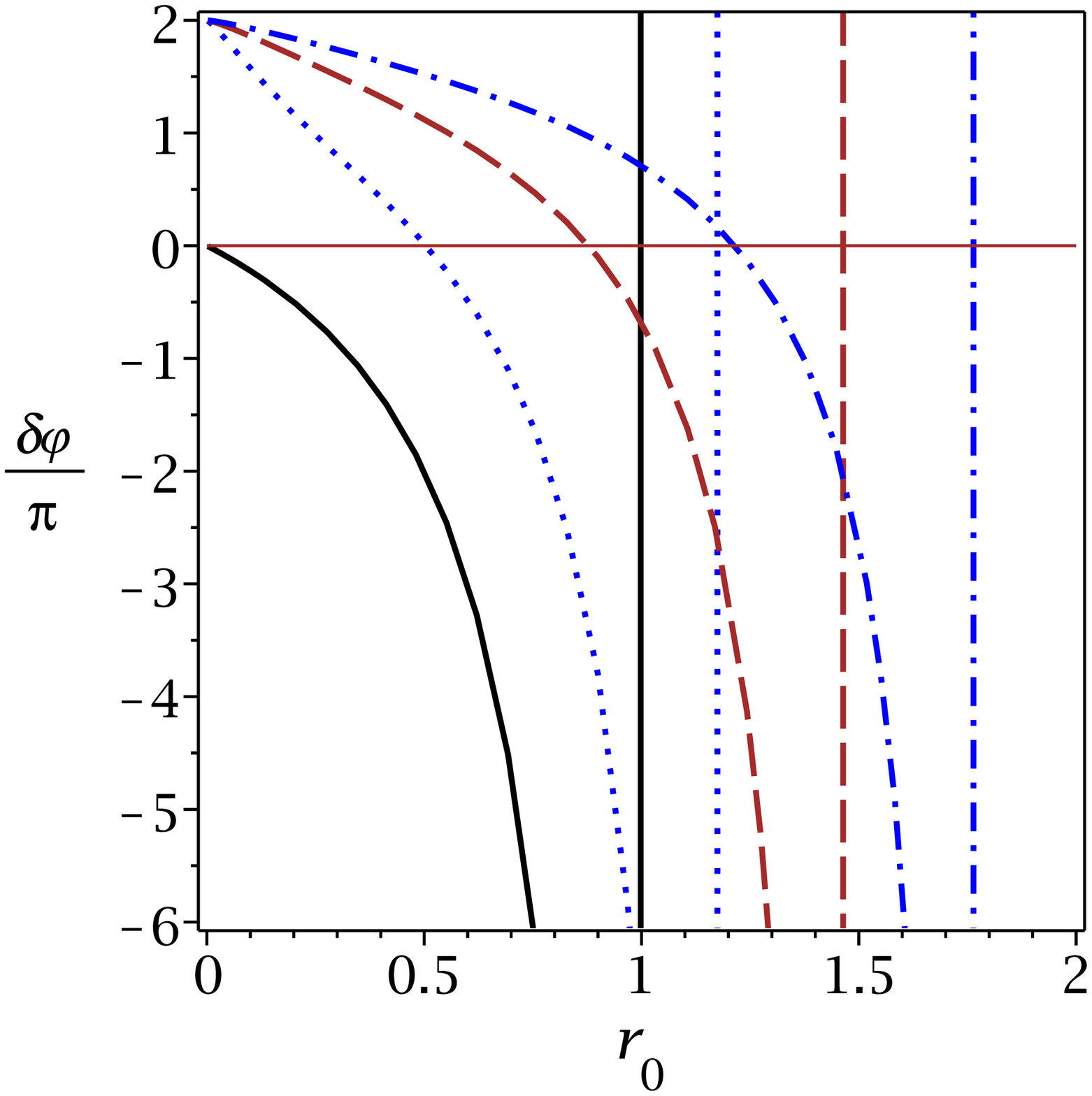}}%
\end{array}
$%
\caption{\textbf{\emph{PMI solutions:}} $\protect\delta \protect\varphi $
versus $r_{0}$ for $m=1$, $c=1$, $c_{1}=2$ and $s=0.9$, $q=0$ (continuous
line), $q=0.5$ (dotted line), $q=1$ (dashed line) and $q=1.5$ (dashed-dotted
line). \newline
\textbf{Left diagram:} $\Lambda =-1$; \textbf{Right diagram:} $\Lambda =1$.}
\label{Fig8}
\end{figure}
\begin{figure}[tbp]
$%
\begin{array}{cc}
\resizebox{0.35\textwidth}{!}{ \includegraphics{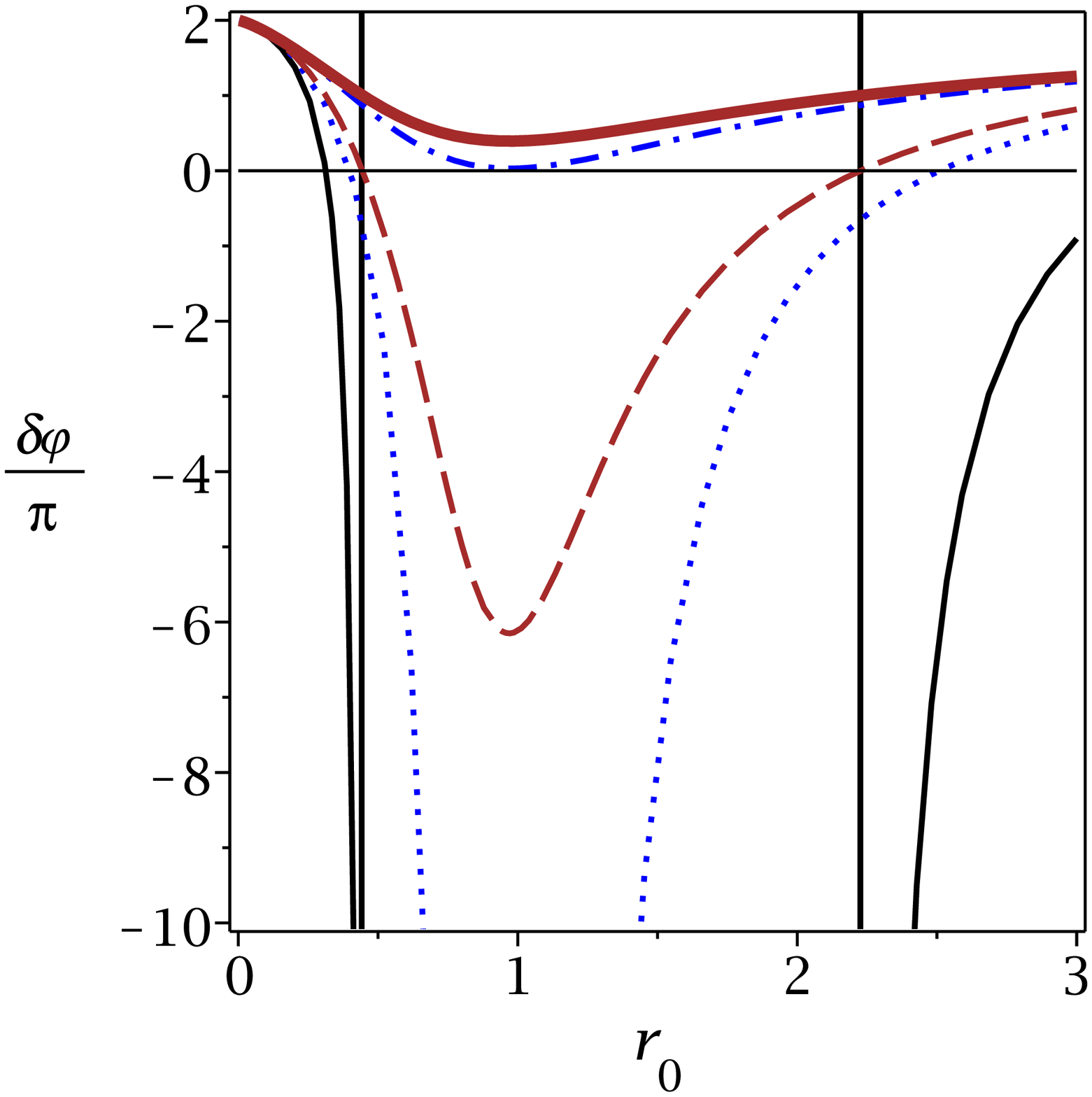}} & %
\resizebox{0.35\textwidth}{!}{ \includegraphics{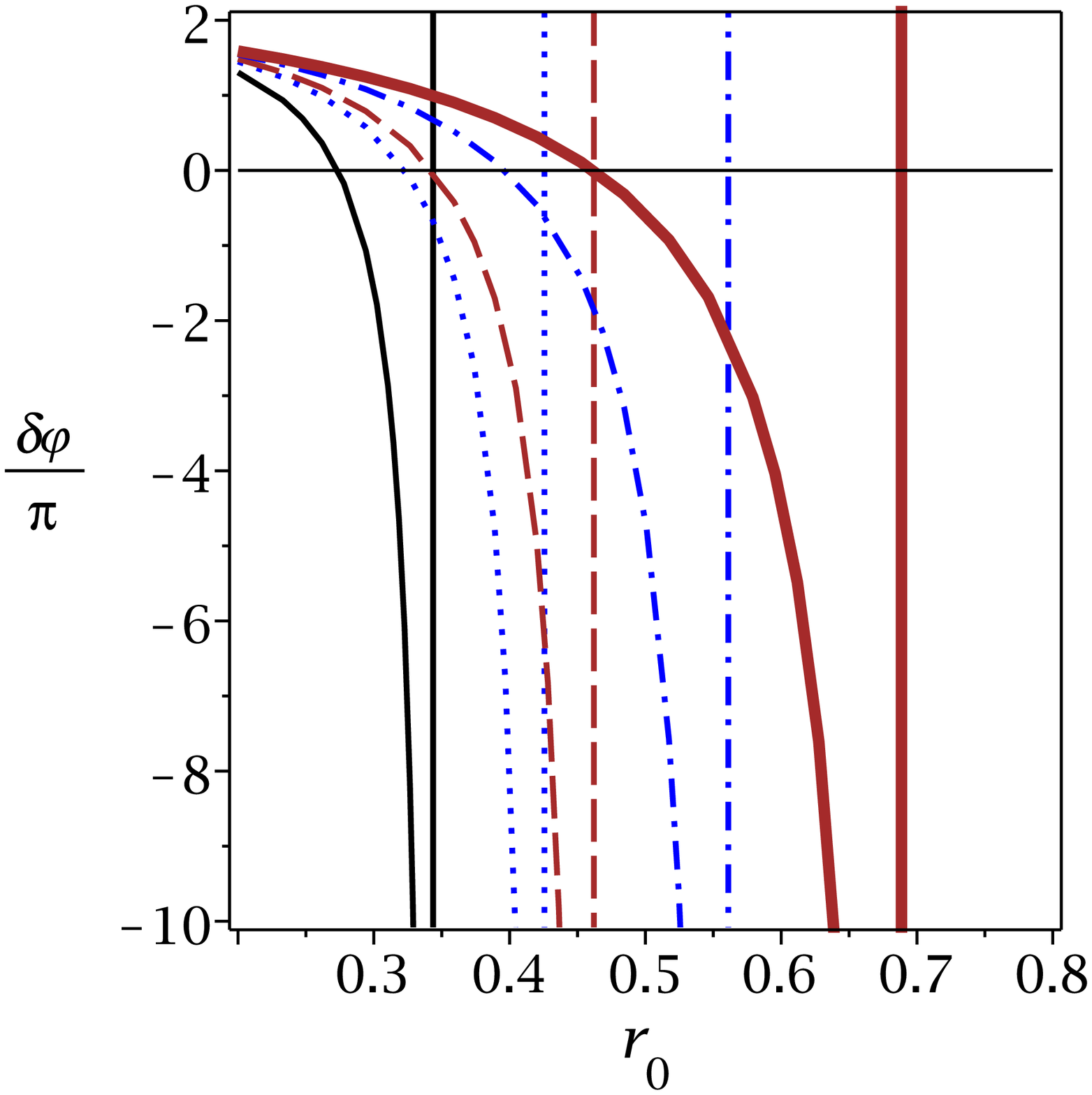}}%
\end{array}
$%
\caption{\textbf{\emph{PMI solutions:}} $\protect\delta \protect\varphi $
versus $r_{0}$ for $q=0.1$, $c=1$, $m=1$ and $s=0.9$, $c_{1}=-5$ (continuous
line), $c_{1}=-3.49$ (dotted line), $c_{1}=-3$ (dashed line), $c_{1}=-1.46$
(dashed-dotted line) and $c_{1}=-1$ (bold line). \newline
\textbf{Left diagram:} $\Lambda =-1$; \textbf{Right diagram:} $\Lambda =1$.}
\label{Fig9}
\end{figure}
\begin{figure}[tbp]
$%
\begin{array}{cc}
\resizebox{0.35\textwidth}{!}{ \includegraphics{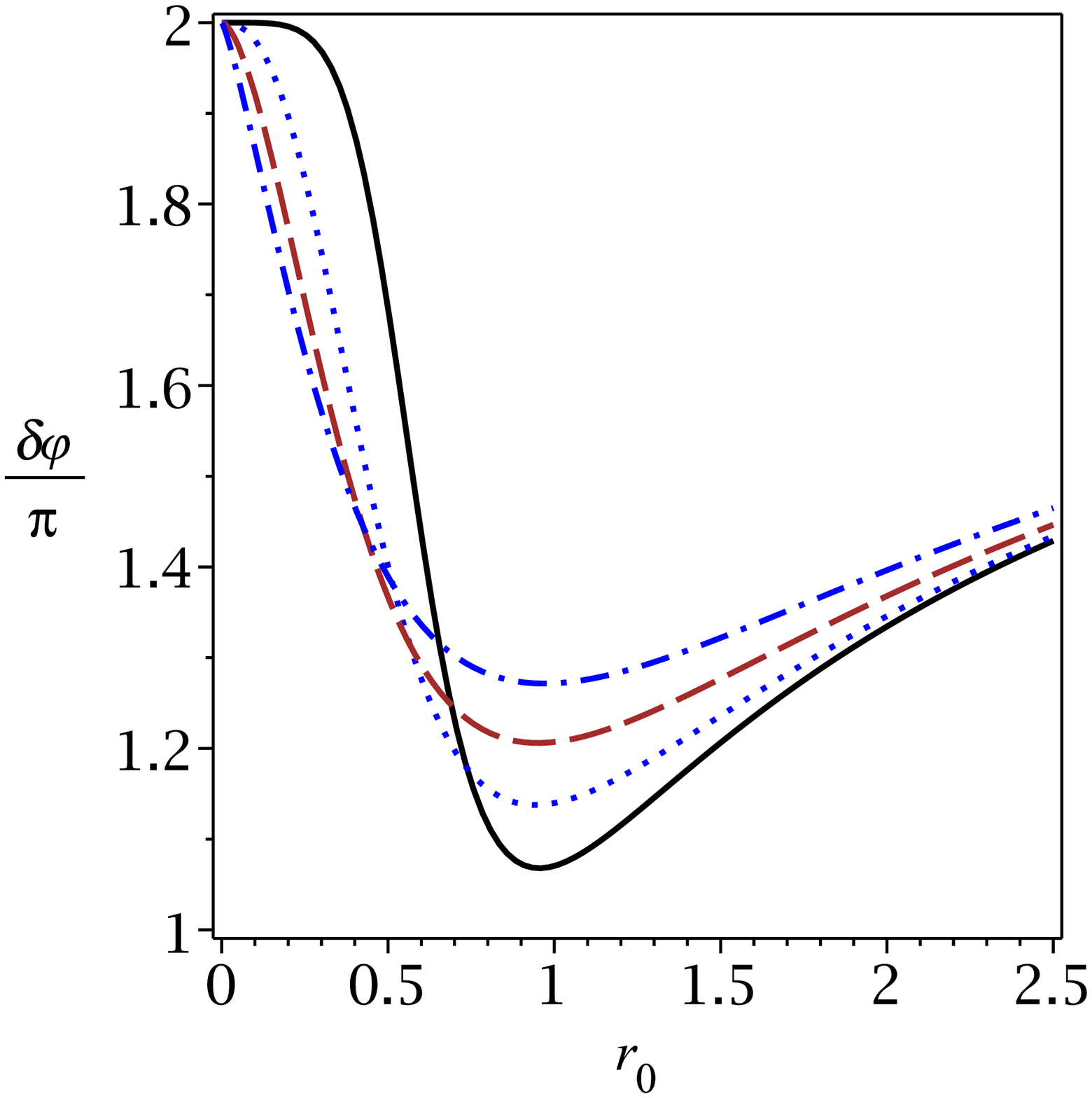}} & %
\resizebox{0.35\textwidth}{!}{ \includegraphics{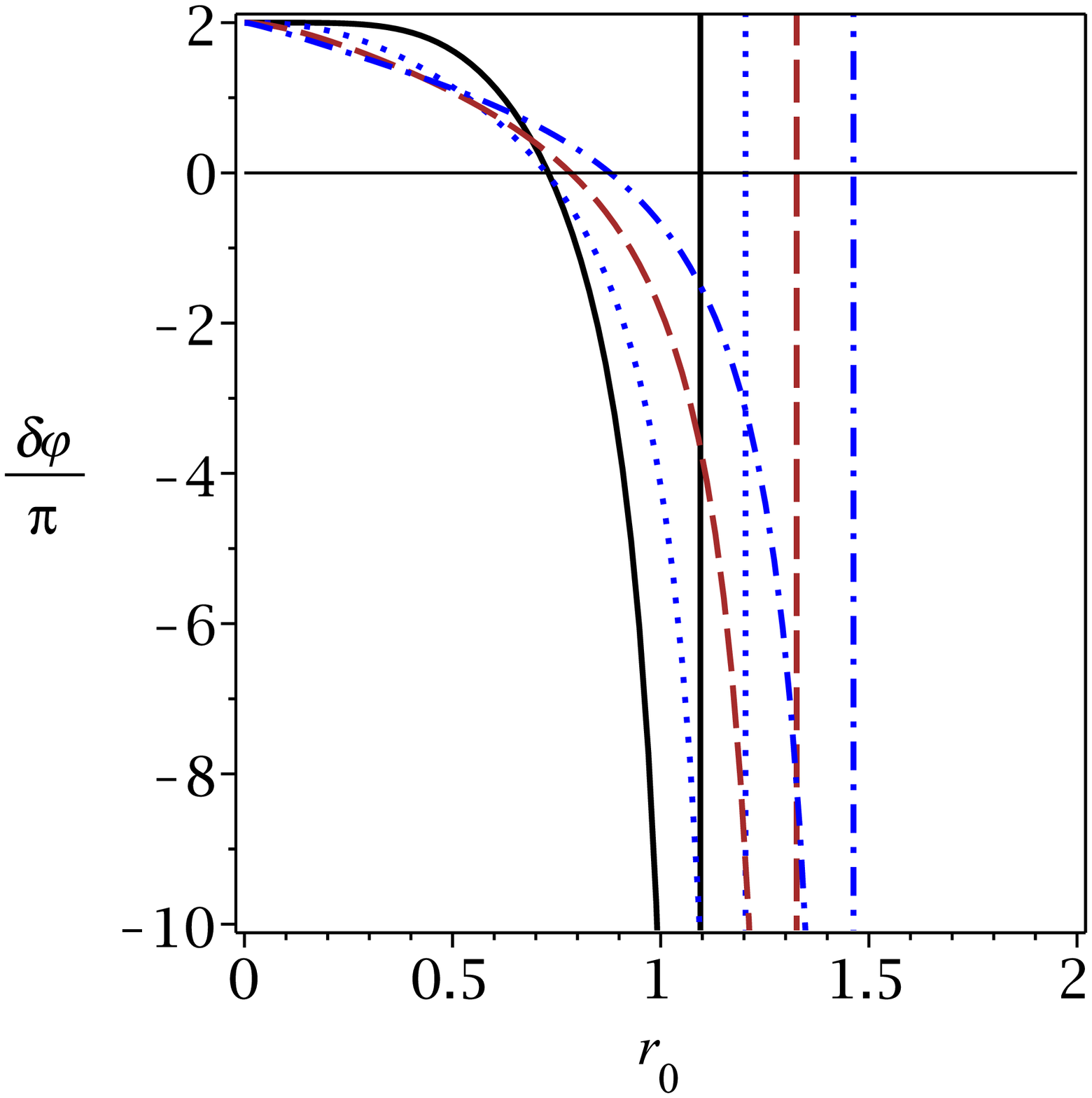}}%
\end{array}
$%
\caption{\textbf{\emph{PMI solutions:}} $\protect\delta \protect\varphi $
versus $r_{0}$ for $q=0.1$, $c=1$, $c_{1}=2$ and $m=1$, $s=0.6$ (continuous
line), $s=0.7$ (dotted line), $s=0.8$ (dashed line) and $s=0.9$
(dashed-dotted line). \newline
\textbf{Left diagram:} $\Lambda =-1$; \textbf{Right diagram:} $\Lambda =1$.}
\label{Fig10}
\end{figure}

\begin{figure}[tbp]
$%
\begin{array}{ccc}
\resizebox{0.35\textwidth}{!}{ \includegraphics{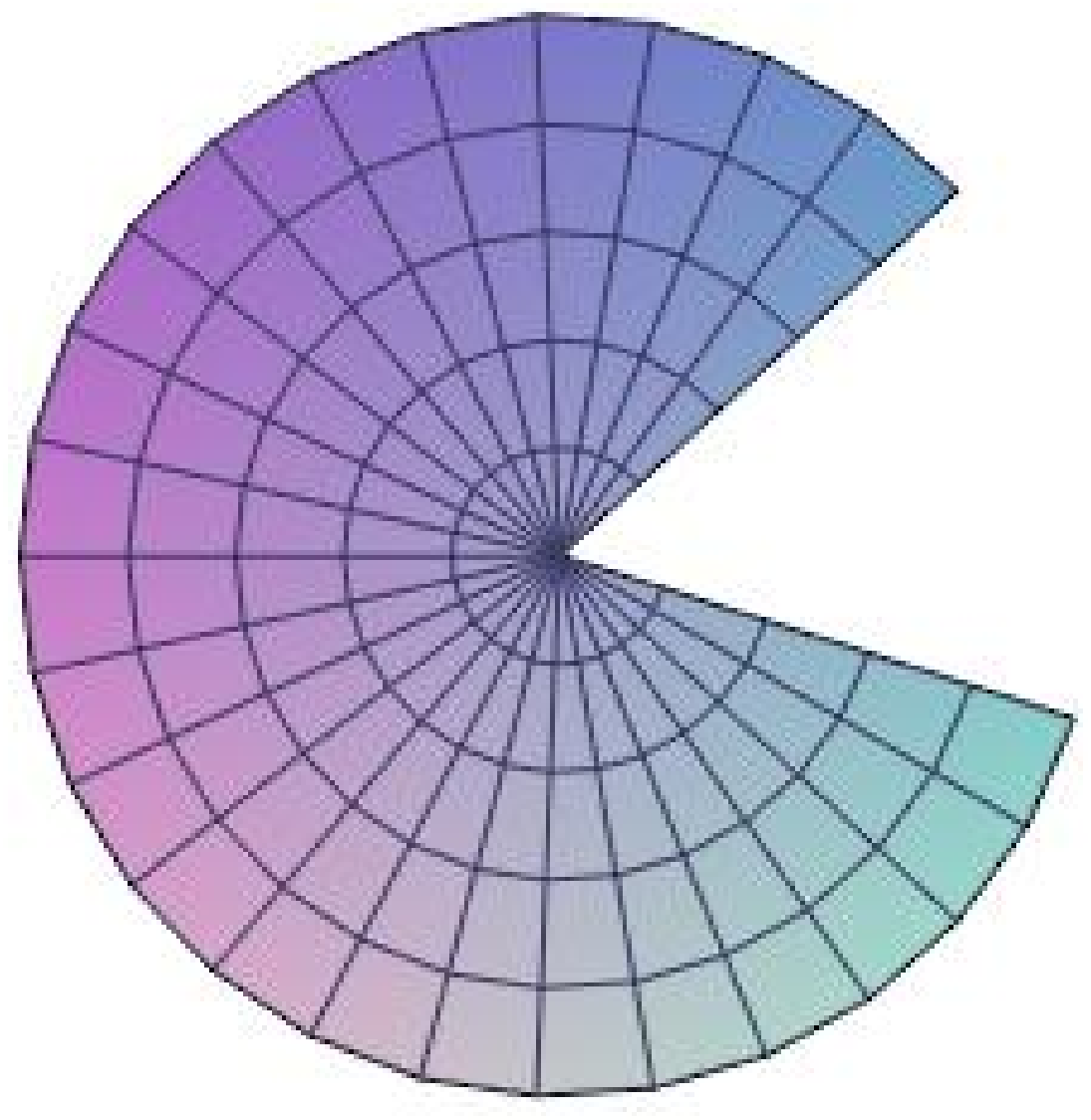}} & %
\resizebox{0.35\textwidth}{!}{ \includegraphics{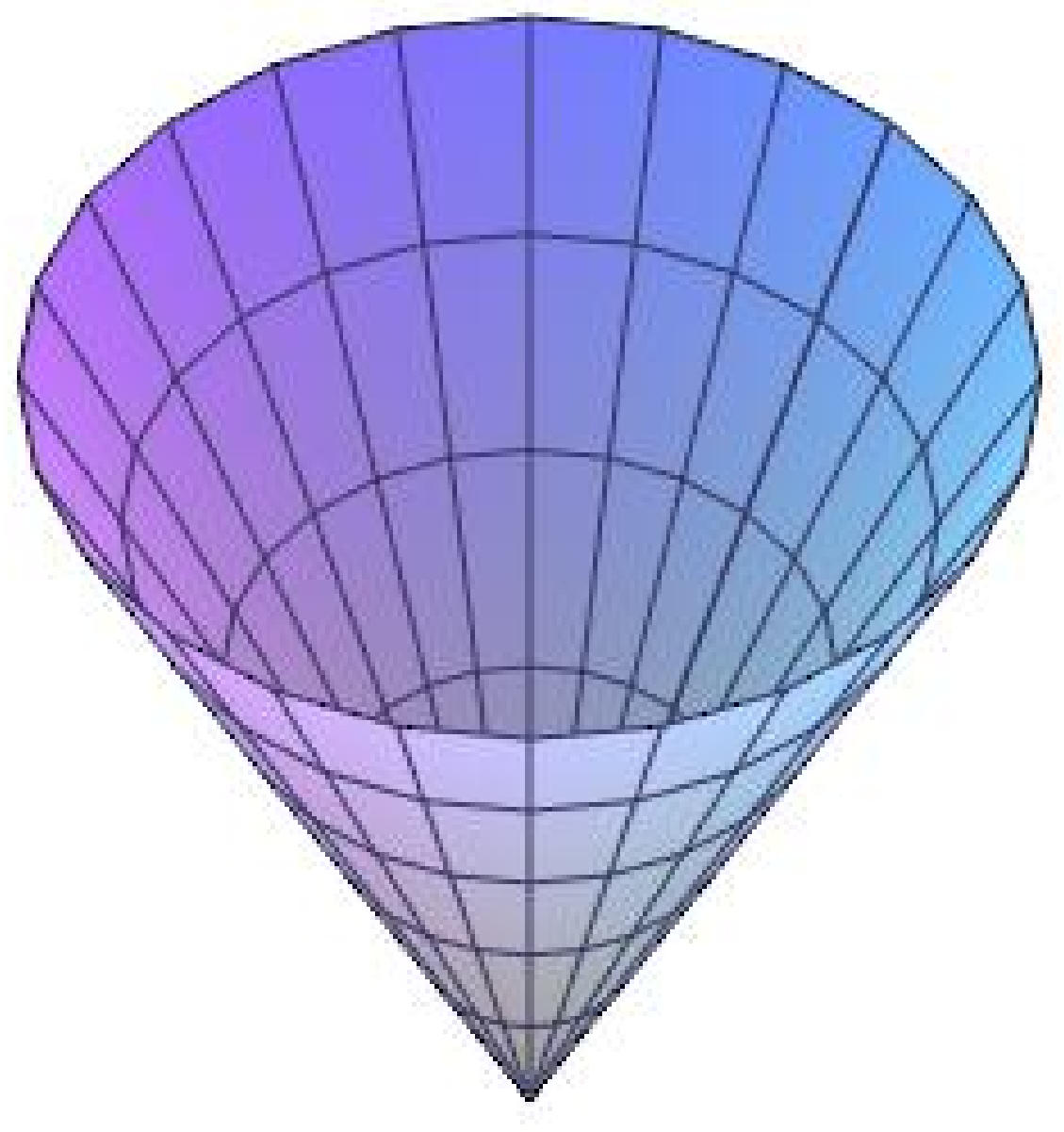}} &
\end{array}
$%
\caption{deficit angle: by sewing the two edges together (left
panel), a cone is formed (right panel).} \label{Figdef}
\end{figure}
\begin{figure}[tbp]
$%
\begin{array}{ccc}
\resizebox{0.35\textwidth}{!}{ \includegraphics{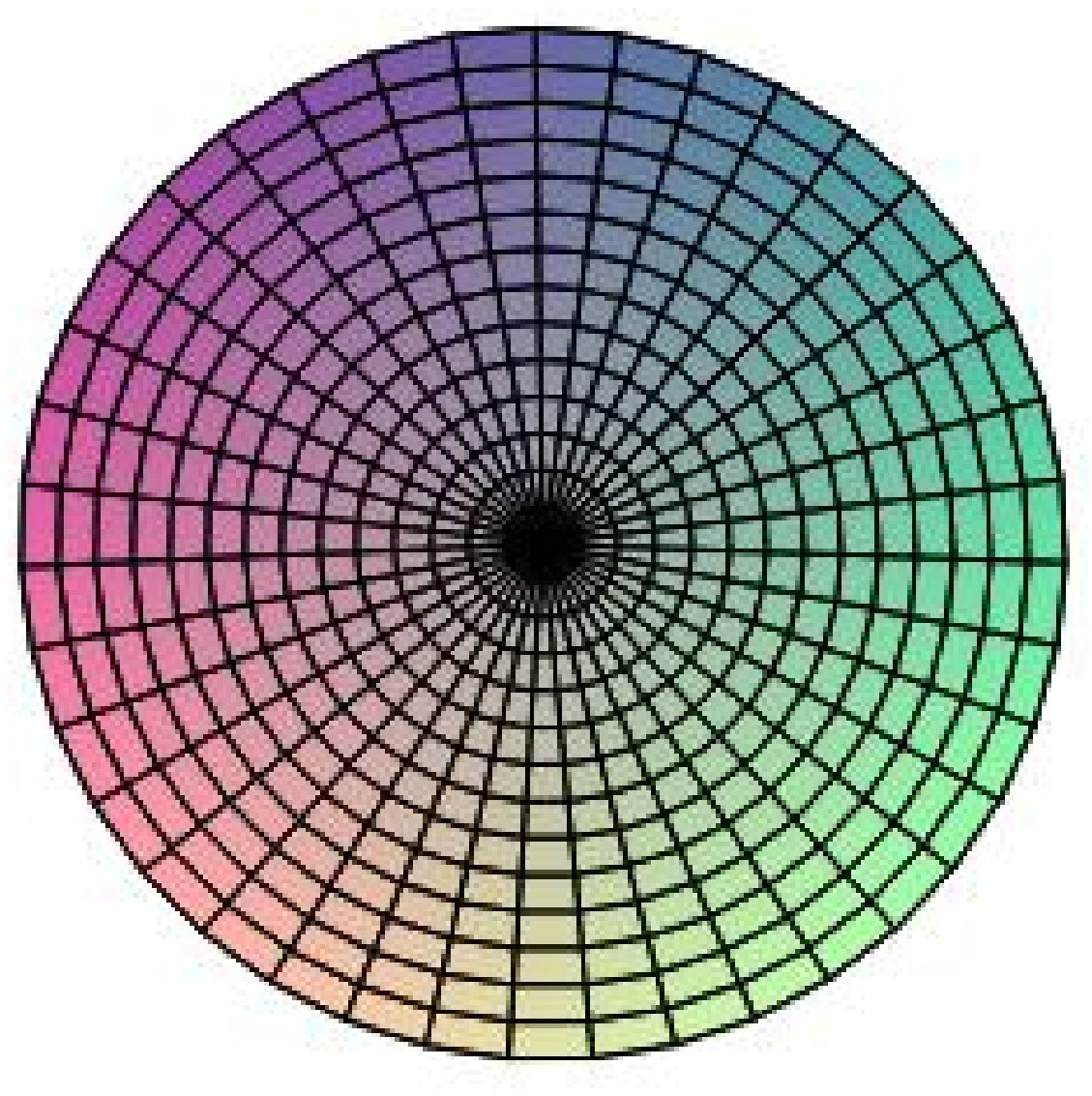}} & %
\resizebox{0.35\textwidth}{!}{ \includegraphics{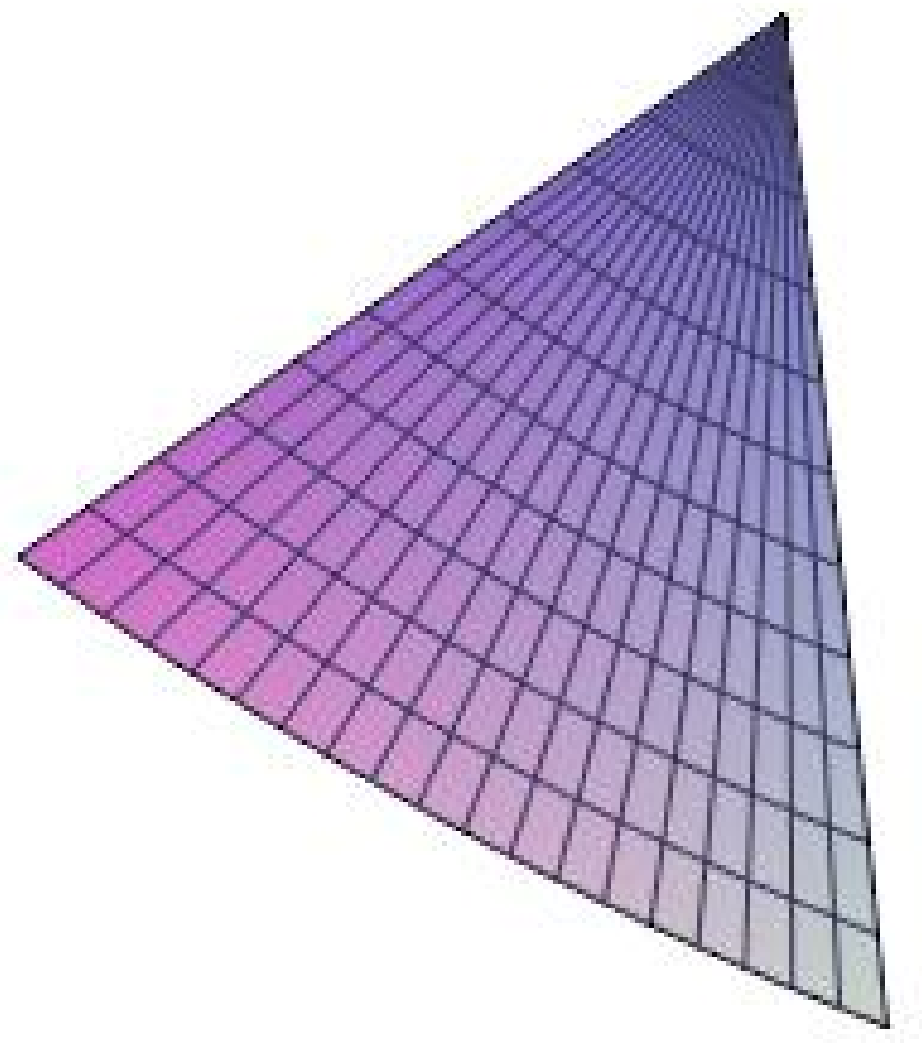}} & %
\resizebox{0.35\textwidth}{!}{ \includegraphics{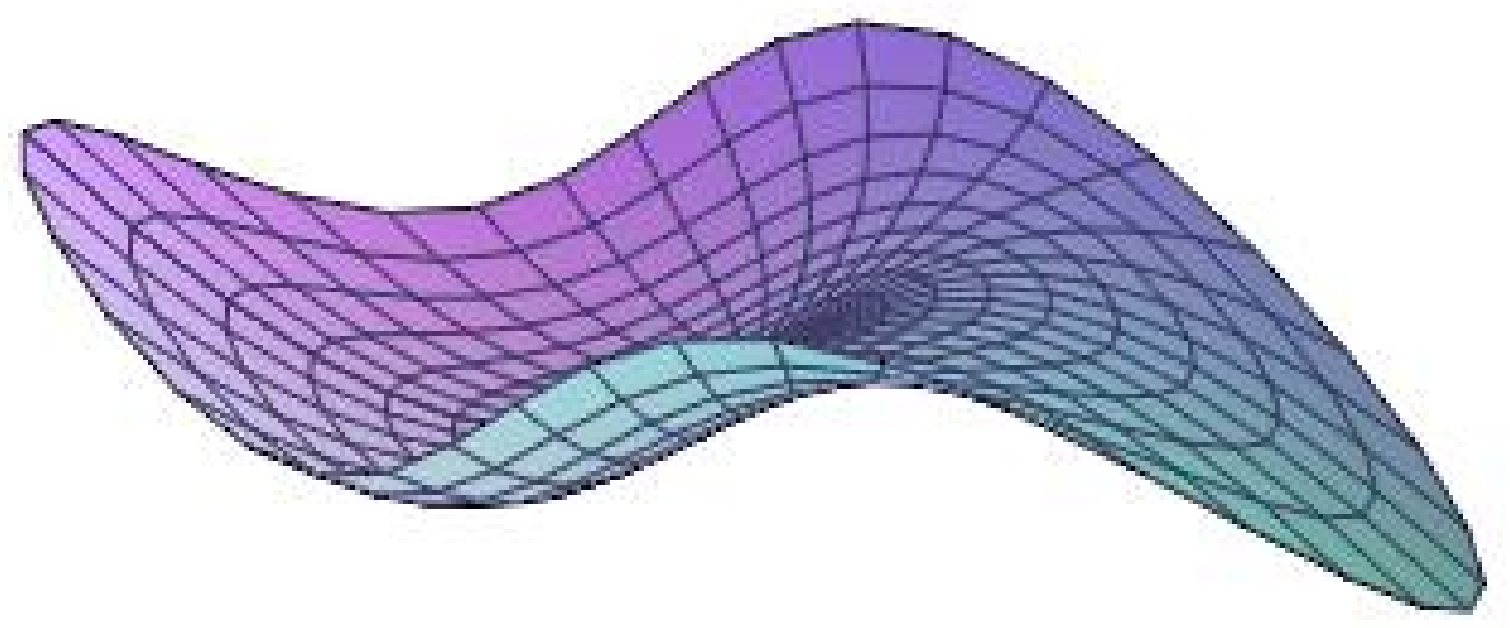}}%
\end{array}
$%
\caption{surplus angle: by adding additional angle (middle panel)
to a circle (left panel), we obtain a new figure (right panel).}
\label{Figsur}
\end{figure}

Depending on the choices of different parameters, the deficit
angle of Maxwell-adS and PMI-adS solutions could have a minimum.
In adS case, except for neutral solutions, the deficit angle could
have; I) Two roots with one region of negativity located between
these two roots. II) One extreme root located at the minimum with
deficit angle being only positive. III) No root and deficit angle
is always positive. The minimum is an increasing function of the
massive parameter (left panels of Figs. \ref{Fig4} and
\ref{Fig7}), electric charge (left panels of Fig. \ref{Fig5} and
\ref{Fig8}) and $c_{1}$ (left panels of Fig. \ref{Fig6} and
\ref{Fig9}). By considering negative values for $c_{1}$, it is
possible to have one of the following cases; I) One divergency
located between two roots. II) Two divergencies which are located
between two roots. Between the divergencies, the deficit angle is
positive but its value is out of the permitted values. In these
two cases, the positive deficit angle could only be observed
before smaller root and after larger root. Interestingly, in the
absence of electric charge, the deficit angle is only an
increasing function of $r_{0}$ (left panels of Figs. \ref{Fig5}
and \ref{Fig8}).

For the Maxwell-dS and PMI-dS spacetimes, interestingly, only one
root and divergency are observed. The root and divergency are
increasing functions of the massive gravity (right panels of Figs.
\ref{Fig4} and \ref{Fig7}), electric charge (right panels of Figs.
\ref{Fig5} and \ref{Fig8}) and $c_{1}$ (right panels of Figs.
\ref{Fig6} and \ref{Fig9}). The divergency is located after root.
The deficit angle is only positive before root. After divergency,
the deficit angle is positive but its values are not in permitted
region. The only exception is for the absence of electric charge
(right panels of Figs. \ref{Fig5} and \ref{Fig8}). In this case,
no root is observed and deficit angle is negative valued.

In the case of PMI theory, another free parameter (nonlinearity parameter)
exists. Evidently, the minimum in adS case is an increasing function of this
parameter (left panel of Fig. \ref{Fig10}). For dS spacetime, the root and
divergency are increasing functions of this parameter.

Depending on values of deficit angle, the geometrical structure of the
magnetic solutions will be determined. Our solutions contain a conical
singularity. This conical singularity is built by considering a $2$%
-dimensional plane replaced with cutting an arbitrary slice and sewing
together the edges. The singular point is located at the apex of cone. Now,
considering this concept, one can see that positive values of the deficit
angle represent missing segment of the $2$-dimensional plane (Fig. \ref%
{Figdef}). On the contrary, the negative values of the deficit angle
represent the additional part that we can add to the mentioned plane (Fig. %
\ref{Figsur}). Therefore, the positivity/negativity of the deficit angle
plays a crucial role in the topological structure of the solutions. Here, we
see that depending on choices of different parameters, it is possible to
obtain negative and positive values of the deficit angle. The roots of
deficit angle could be interpreted as transition points in which the total
shape of the object is modified. On the other hand, the existence of
divergencies for deficit angle marks the possibility of the absence of
magnetic solutions which was observed for both the dS and adS spacetimes.
Previously, through several studies, it was shown that existence of
deficit/surplus angle enables one to regard the cosmological constant
problem \cite{def1}. The main motivation of this paper was understanding the
effects of massive gravity and PMI theory on the magnetic solutions. The
variation in deficit angle shows that the total structure of the solutions
depends on contributions of these two generalizations. Specially, we
observed that generalization to massive gravity provided the possibility of
existence of divergence points for adS spacetime. It is worthwhile to
mention that for adS case, between two divergencies, the values of deficit
angle are within prohibited range. This indicates that there is no
acceptable deficit angle between the divergencies in adS case.

\section{Conclusions}

In this paper, we have considered magnetic solutions which contain a conical
singularity without any event horizon and curvature singularity. The set up
for the gravity and energy momentum tensor were consideration of two
generalizations: massive gravity and PMI nonlinear electromagnetic field.

The geometrical properties of the solutions were obtained and
deficit angle for the two cases of Maxwell-massive and PMI-massive
were extracted. It was shown that the general structure of the
solutions depends on choices of different parameters through
positivity and negativity of the deficit angle. Existence of root
and divergency were reported and it was shown that these
properties of the solutions depend on the choices of different
parameters, such as massive gravity and nonlinearity parameter. In
addition, it was shown that depending on the nature of background
(being dS or adS), deficit angle, hence geometrical structure of
the solutions would be different. The difference was highlighted
analytically and numerically through several diagrams.

The existence of root and divergency for deficit angle was reported which
indicates that under certain conditions, suitable choices of different
parameters, topological defects known as magnetic solutions would enjoy
geometrical phase transition. The dependency of geometrical phase transition
on nonlinearity parameter and massive gravity highlighted the importance and
roles of massive gravity and also nonlinear electromagnetic field
generalizations. Especially, the existence of divergency for adS spacetime
in the presence of massive gravity could be pointed out.

The existence of deficit and surplus angles results into two
completely different astrophysical objects which essentially
requires different methods for detection (see Figs. \ref{Figdef}
and \ref{Figsur}). In fact, when we are talking about deficit
angle, it means that the geometrical structure of the solutions
enjoys a positive tension in their structures. On the contrary,
existence of the surplus angle corresponds to presence of the
negative tension \cite{de Rham3}. In this paper, we showed that
depending on choices of different parameter, the possibility of
both are provided for our magnetic solutions. In fact, in some
cases, the existence of discontinuity, hence phase transition
between deficit angle and surplus angle was reported for our
solutions. Considering the important applications of the
deficit/surplus angle in the context of cosmology and cosmological
constant problem, one can employ the results of present paper to
understand the roles of massive gravity and nonlinear
electromagnetic field on these applications and their
corresponding results. We leave these matters for future works.

\begin{acknowledgements}
The authors wish to thank Shiraz University Research Council. This
work has been supported financially by Research Institute for
Astronomy and Astrophysics of Maragha.
\end{acknowledgements}

\end{document}